\documentclass[10pt]{extarticle}
\usepackage{fullpage}
\usepackage[english]{babel}
\usepackage[utf8]{inputenc}
\usepackage{amsmath,amssymb,mathtools,xy,lscape} 
\usepackage{amsmath, amssymb, bm}
\usepackage{graphicx}
\usepackage{subcaption}
\usepackage[colorinlistoftodos]{todonotes}
\usepackage{indentfirst}
\xyoption{all}    

\newcommand{\hiddenpower}[2] { \ifnum \numexpr#2=1 #1 \else #1^#2 \fi }
\numberwithin{equation}{section}

\usepackage{xparse}
\ExplSyntaxOn
\newcounter{diff_order}
\newcounter{diff_power}

\newcommand{\rawdiff}[3]
{
	\setcounter{diff_order}{0}
	\clist_map_inline:nn{#3}{\stepcounter{diff_order}}
	
	\frac{\hiddenpower{#1}{\thediff_order} #2}
	{
		\def\old_var{DefaultValue}
		\setcounter{diff_power}{0}
		
		\clist_map_inline:nn{#3}
		{
			\def\new_var{##1}
			\ifnum \thediff_power=0
				\stepcounter{diff_power}
			\else
				\tl_if_eq:NNTF \new_var \old_var
				{\stepcounter{diff_power}}
				{
					#1 \hiddenpower{\old_var}{\thediff_power}
					\setcounter{diff_power}{1}
				}
			\fi

			\def\old_var{##1}
		}
		
		#1 \hiddenpower{\old_var}{\thediff_power}
	}
}

\newlength{\bibitemsep}
\newlength{\bibparskip}
\let\oldthebibliography\thebibliography
\renewcommand\thebibliography[1]{%
  \oldthebibliography{#1}%
  \setlength{\parskip}{\bibitemsep}%
  \setlength{\itemsep}{\bibparskip}%
}

\ExplSyntaxOff

\newcommand{\lb}{\left(}
\newcommand{\rb}{\right)}

\renewcommand{\sinh}[2][1]{\hiddenpower{\text{sinh}}{#1} \lb #2 \rb}
\renewcommand{\cosh}[2][1]{\hiddenpower{\text{cosh}}{#1} \lb #2 \rb}

\renewcommand{\ln}[1]{\text{ln} \lb #1 \rb}

\begin{document}


\begin{center}
\strut\hfill


\noindent { {\bf {DISCRETIZATIONS OF THE GENERALIZED AKNS SCHEME}}}\\
\vskip 0.35in

\noindent {{  {\footnotesize ANASTASIA DOIKOU AND SPYRIDOULA SKLAVENITI}}}
\vskip 0.25in

\noindent {\footnotesize School of Mathematical and Computer Sciences, Heriot-Watt University,\\
Edinburgh EH14 4AS, United Kingdom}

\vskip 0.15in

\noindent {\footnotesize {\tt E-mail: a.doikou@hw.ac.uk, ss153@hw.ac.uk}}\\

\vskip 0.80in

\end{center}

\begin{abstract}
\noindent We consider space discretizations of the matrix Zakharov-Shabat AKNS scheme, in particular
the discrete matrix non-linear Scrhr\"odinger (DNLS) model,
and the matrix generalization of the Ablowitz-Ladik (AL) model, 
which is the more widely acknowledged discretization.  We focus on the derivation of solutions via local 
Darboux transforms for both discretizations, and we derive novel solutions 
via generic solutions of the associated discrete  linear equations. 
The continuum analogue is also discussed, and as an example 
we identify solutions of
the matrix NLS equation  in terms of the heat kernel.
In this frame we also derive a discretization 
of the Burgers equation via the analogue of the Cole-Hopf transform.
Using the basic Darboux transforms for each scheme we identify  both matrix 
DNLS-like and AL hierarchies, i.e. we extract the associated Lax pairs, via the dressing process.
We  also discuss the global  Darboux transform, which is the discrete analogue of the integral
transform, through the discrete Gelfand-Levitan-Marchenko (GLM) equation. 
The derivation of the discrete matrix GLM equation and associated solutions 
are also presented together with explicit linearizations. 
Particular emphasis is given in the discretization schemes, i.e. forward/backward  in the 
discrete matrix DNLS scheme versus symmetric in the discrete matrix AL model.
\end{abstract}




\section{Introduction}

\noindent 
Non-linear Schr\"odinger-type  systems (AKNS scheme) are the most well studied integrable hierarchies 
(see for instance \cite{AKNS1}--\cite{DoFiSk1} and references therein). 
Both continuum and discrete versions \cite{AL}--\cite{Sklyanin} 
 have  been widely studied from the point of view of the inverse scattering 
method or the Darboux and Zakharov-Shabat (ZS) dressing methods 
\cite{ZakharovShabat}--\cite{DoFiSk1}, \cite{Chud}--\cite{Rourke}, yielding
solutions of hierarchies of integrable non-linear PDEs (ODEs) as well as  hierarchies of  associated Lax pairs. 
In this investigation we are focusing on space discretizations of the NLS-type hierarchy.
Specifically we consider two versions of the generalized semi-discrete NLS-type hierarchies: 
1) the discrete system introduced in \cite{KunduRagnisco, Sklyanin} (DNLS), which is the natural discretization of 
the NLS model (AKNS more generally), as will be transparent below. 
2) The generalized AL model (see e.g. \cite{Ablo,  AL, Kulish}), 
which is the widely studied version,  although not the most natural discretization from the algebraic point of view. 

More precisely, our main aim is the study of the 
$\mbox{G}r({\cal N}| {\cal N}+ {\cal M})$ Grassmannian extension of the DNLS hierarchy and the AL model 
(see \cite{KunduRagnisco, Sklyanin} and \cite{AL, Kulish}, \cite{Chud}--\cite{Rourke}, see also \cite{Ablo} 
for a detailed list of references on NLS and AL models as well as  
their multi-component generalizations).
Generalized local \cite{Darboux} and global Darboux type \cite{ZakharovShabat, Ablo} transformations are
 employed in order to identify solutions 
of the associated hierarchies of the  nonlinear ODEs as well as produce the semi-discrete hierarchies, 
(regarding the continuous generalizations we refer the interested reader to \cite{Ablo, Degalomba1, DoFiSk1}). 
The proposed DNLS-type hierarchy, we are most interested in in this study, provides in fact a discretization scheme 
of the matrix AKNS scheme \cite{AKNS1};  indeed discrete generalized 
versions of the mKdV and mKP naturally arise in this frame. It is also worth noting that  to our knowledge
this is the first time that generic solutions 
for the DNLS hierarchy are systematically derived, as the majority of the studies on the discretization of AKNS 
refer to the AL model.
Note also that the DNLS hierarchy (see \cite{Sklyanin} and references therein) is a natural integrable 
version of the discrete-self-trapping (DST)  equation introduced and studied in \cite{DST} 
to model the nonlinear dynamics of small molecules, such as ammonia,
acetylene, benzene, as well as large molecules, such as acetanilide. It is also related to
various physical problems such as arrays of coupled nonlinear
wave-guides in nonlinear optics and quasi-particle motion on a dimer among others. 

The fundamental object in our approach is the Lax operator, i.e. given the Lax operator and implementing 
the dressing method we will 
be able to identify all the members of the two distinct discrete hierarchies. 
 We should note that numerous  studies from the Hamiltonian point of view \cite{FT} in the case of periodic as well as 
generic integrable boundary conditions (see for instance \cite{Sklyanin2, DoFiSk2}) exist. 
The algebraic description allows the computation of the time components of the Lax pairs of the hierarchy 
via the fundamental Semenov-Tian-Shansky formula \cite{STS}, that involves the $r$ and $L$ matrices. 
This universal formula has been extended  in the case of open boundary conditions  \cite{AvanDoikou1}, 
as well as at the quantum level  \cite{DoikouFindlay1}. 
Algebraically speaking the Lax operators of the two schemes, 
DNLS and AL, are representations of two distinct deformed algebras: 
the DNLS model is associated to the classical Yangian (rational) $r$-matrix similarly to 
the continuous NLS model, while the AL model is associated to 
a trigonometric $r$-matrix. 
Similarly, at the quantum level DNLS  corresponds to coupled quantum harmonic oscillators and is 
a representation of the Yangian \cite{Sklyanin}, whereas the 
quantum AL or $q$-boson model \cite{Kulish, Korff} corresponds to deformed harmonic oscillators and
is associated to the $\mathfrak{U}_q(\mathfrak{sl}_2)$ $R$ matrix \cite{Jimbo}.

Let us focus here on the classical case, and discuss in more detail the Hamiltonian/algebraic  description for both 
the {\it scalar} DNLS and AL models, and illustrate how the DNLS scheme is the more natural discretization 
from the algebraic viewpoint compared to the AL one.
The Lax operator for both models satisfies Sklyanin's bracket \cite{FT}:
\begin{equation}
\Big \{ L(\lambda) \underset{^{,}}\otimes L(\mu)\Big \} = \Big [r(\lambda- \mu),\ L(\lambda) \otimes L(\mu) \Big ], \label{Poisson}
\end{equation}
$\lambda,\ \mu$ are spectral parameters, $r$ is the classical matrix that satisfies the classical Yang-Baxter equation. 
The $r$-matrix for the classical scalar DNLS model is the $\mathfrak{sl}_2$ Yangian matrix.
In general, the $\mathfrak{gl}_{{\mathrm k}}$ Yangian $r$-matrix \cite{Yang} is given as
\begin{equation}
r(\lambda) = {1\over \lambda}\sum_{i,j =1}^{{\mathrm k}}  e_{ij}\otimes e_{ji}
\end{equation}
where $e_{ij}$ are ${\mathrm k} \times {\mathrm k}$ matrices with entries: $(e_{ij})_{kl} = \delta_{ik} \delta_{jl}$.
The $r$-matrix for AL model on the other hand is a trigonometric matrix,  a variation of the classical sine Gordon $r$-matrix \cite{Kulish, FT}:
\begin{equation}
r(\lambda) = {1\over \sinh \lambda } \Big (\cosh\lambda  \sum_{j=1}^2 e_{jj} \otimes e_{jj}  +
 \sinh \lambda \sum_{i \neq j=1}^2 e_{ij} \otimes e_{ji} + \sum_{i\neq j=1}^2 (-1)^{j-i} e_{ii} \otimes e_{jj} \Big).
\end{equation}
We also recall that the classical Lax operators for the scalar DNLS and AL models are given by (see e.g. \cite{AL, Sklyanin, Kono})
\begin{eqnarray}
&& \mbox{DNLS model:} ~~~~ L_n(\lambda) = \lb \begin{matrix}
		\lambda +1+x_n y_n & x_n\\
		y_n & 1
	\end{matrix} \rb \label{DNLS00}\\
&& \mbox{AL model:} ~~~~~ L_n(z) = \lb \begin{matrix}
		z & \hat \beta_n\\
		\beta_n &  z^{-1}
	\end{matrix} \rb, \label{AL00}
\end{eqnarray}
where $z =e^{\lambda}$ is the multiplicative spectral parameter.

The Poisson structure (\ref{Poisson}) leads to two distinct algebras at the classical level \cite{Sklyanin, KakMug}:
\begin{eqnarray}
&&\mbox{DNLS:} ~~~ \Big\{ x_n,\ y_m \Big \} =-\delta_{nm}, ~~~\Big \{x_n, \ x_m \Big\} = \Big \{y_n, \ y_m \Big\} =0 \label{dn1} \\
&& \mbox{AL:} ~~~ \Big \{\beta_n,\ \hat \beta_m \Big\}= \delta_{nm} \big ( 1 - \beta_n\hat \beta_m\big ), ~~~
\Big \{\beta_n,\ \beta_m  \Big\}= \Big\{\hat \beta_n,\ \hat \beta_m\Big \}=0.
\end{eqnarray}
The AL model is thus  associated to a deformed harmonic oscillator  classical 
algebra ($q$-bosons at the quantum level). 
The generalized DNLS-type hierarchy is associated to a canonical algebra i.e. the harmonic oscillator model and is immediately 
associated to the NLS model for which the corresponding exchange relations satisfied by the continuous  fields $\hat u,\ u$ 
are the continuous analogues of (\ref{dn1}): 
$ \Big\{ \hat u(x),\  u(y) \Big \} =-\delta(x,y)$.

It is worth noting that the Hamiltonian/algebraic description offers the strongest statement of integrability at both classical and quantum level, 
given that it naturally provides all the charges in involution. However, when deriving solutions of the associated 
integrable PDEs (ODEs) it is more efficient to consider the Lax pair picture and the dressing schemes,  and this is precisely 
the formulation that we are adopting here.

Let us briefly outline what is achieved in the article:
\begin{itemize} 
\item  In section 2 we introduce the Grassmannian DNLS Lax operator as well
as two fundamental Darboux matrices. Given the $L$-matrix and the local Darboux transforms \cite{Darboux}, we compute 
the associated integrals of motion, and we also derive certain solution of the corresponding non-linear ODEs given the specific choice 
of the Darboux transform.  Employing one of the fundamental Darboux matrices we also perform the dressing process and we identify 
the Lax pairs of the hierarchy; 
explicit expressions for the first few members are presented (up to the Hamiltonian and the Lax pair for the generalized matrix mKdV equation).  
In the Appendix the recursion formula for the conserved quantities is presented together with the details on the Lax pair of the generalized mKdV equation.
Also, the Poisson structure for the associated fields is extracted by comparing the equations of motion emerging  from the zero curvature condition
and from  the Hamiltonians.
We derive two types of solitonic solutions, and we also produce explicit analytic expressions for
the two-soliton solution using Bianchi's permutability theorem. More importantly, with the use of a Toda type 
Darboux matrix we identify generic new solutions (i.e. not only solitonic) of the non-linear ODEs in terms of solutions of the associated linear equations.
Such general solutions are also discussed in the continuous case, and for the NLS case in particular we show that they are expressed 
in terms of the heat kernel.
In this context we also derive a discrete version of the viscous Burgers equation via the discrete analogue of the Cole-Hopf transform.

\item In section 3 we implement the dressing scheme for the Grassmannian AL model and we extract the Lax pairs for the discretization of NLS 
as well as solutions via certain local Darboux transforms (see also e.g. \cite{Zullo} and references therein). We basically review some already 
known results, especially regarding the form of the local Darboux transforms, but also compare with the discretization scheme and the results of section 2.
However, in this case too, we are able to derive general solutions
of the non-linear ODEs in terms of solutions of the associated linear equations, via a local Darboux transform, 
and this is the most interesting new result in this section. We also derive the  matrix generalization of the non-linear 
network equations (see e.g.  \cite{Darboux}).

\item In section 4  we introduce the discrete Gelfand-Levitan-Marchenko equation 
(see also \cite{Ablo, Darboux}) emerging from the upper/lower Borel matrix decomposition. 
We also introduce the linearizations of both DNLS and AL systems  via a forward/backward discretization scheme associated to 
DNLS, and  a symmetric scheme associated to AL. In the AL case in the symmetric discretization scheme, 
analogous results can be found in \cite{Ablo} via the Inverse Scattering Transform,
however the DNLS scheme we propose in this section i.e. the forward/backward discretization 
and linearization scheme is new as far as we know.
The corresponding discrete calculus and dispersion relations are also derived. 
General solutions of the discrete GLM  equation in terms of solutions of the corresponding linear equations are then easily extracted. 

\item 
In the last section a general discussion about the main findings of this investigation 
is presented and some open fundamental questions for future studies are also addressed.

\end{itemize}

\section{ The discrete $\mbox{Gr}({\cal N}| {\cal N}+{\cal M})$ NLS-type hierarchy}

\noindent The main aims in this section are: 1) the construction of the integrals of motion 
for the semi-discrete DNLS type hierarchy from the power series expansion of the monodromy  matrix; 2) the derivation of the associated Lax pairs
for the whole hierarchy via the dressing Darboux-B\"acklund scheme and 3) the identification of solutions of the emerging integrable non-linear 
ODEs via suitable choices of Darboux  transformations. In addition to solitonic and multi-solitonic solutions 
we provide general solutions of the non-linear ODEs in terms of solutions of the associated linear equations, expressed as Fourier transforms. 
Such general solutions are also discussed in the continuous case, and for the NLS case in particular we show that they are expressed in 
terms of the heat kernel. Also, the Poisson structure for the associated fields is extracted by comparing the equations of motion from the 
zero curvature condition 
and the associated Hamiltonians.

\subsection{The Lax pair $\&$ the Darboux-B\"acklund transform}

\noindent  The Lax pair in semi-discrete models $(L,\ V)$ depends on the fields (classical or quantum) and some spectral parameter 
and satisfies the auxiliary linear problem 
relations
\begin{eqnarray}
&&\Psi_{n+1} =L_n \Psi_n  \label{aux1}\\
&& \partial_t\Psi_n = V_n \Psi_n, \label{aux2}
\end{eqnarray}
which lead to the discrete equations of motion:
\begin{equation}
\partial_{t_{\alpha}} L_n = V^{(\alpha)}_{n+1}L_n - L_n V^{(\alpha)}_n,
\end{equation}
where the notation above denotes that the Lax pair $(L,\ V^{(\alpha)})$ corresponds to the $t_{\alpha}$ flow.
 
Our main input is the $L$-operator for the non-commutative version of the discrete NLS model (\ref{DNLS00}), 
\begin{equation}
 L_n(\lambda) = \lb \begin{matrix}
		\lambda {\mathbb I}^{({\cal N})} + {\mathbb N}_n & X_n\\
		Y_n & {\mathbb I}^{({\cal M})}
	\end{matrix} \rb, \label{DNLS}
\end{equation}
where $X$ is  an ${\cal N} \times {\cal M}$ matrix, $Y$ an
${\cal M}\times {\cal N}$ matrix, and ${\mathbb N} = \theta {\mathbb I}^{(\cal N)}+ XY $ is an ${\cal N} \times {\cal N}$ matrix, 
$\theta$ is an arbitrary constant.
Our aim now is to identify solutions for the generalized discrete NLS model by means of suitable Darboux transformations.
Let $M_n$ be the local Darboux transform such that 
\begin{equation}
\Psi_n = {\mathbb M}_n \hat \Psi_n, \label{gauge}
\end{equation}
both $\Psi,\ \hat \Psi$ satisfy 
the auxiliary linear problem (\ref{aux1}), (\ref{aux2}) with $(L,\ V)$ and $(\hat L,\ \hat V)$ respectively, then it follows:
\begin{equation}
{\mathbb M}_{n+1}(\lambda)\  \hat L_n(\lambda) = L_n(\lambda)\ {\mathbb M}_n(\lambda), \label{BT1}
\end{equation}
where $\hat L$  is of the same form as $L$ but with fields $\hat X,\ \hat Y$. Equation (\ref{BT1}) can also 
be interpreted as a zero curvature condition in the frame of discrete space-time systems where
$L$ can be thought of as the function at some discrete time $\alpha $ and $\hat L$ as the 
function at $\alpha+1$, so the latter represents a discrete time map.  Similarly, for the time components of the 
Lax pair the transformation (\ref{gauge}) leads to
\begin{equation}
\partial_{t_{\alpha}} {\mathbb M}_n = V^{(\alpha)}_n {\mathbb M}_n- {\mathbb M}_n \hat V^{(\alpha)}_n.  \label{BT2}
\end{equation}
Having at our disposal the two fundamental Darboux-BT equations we can solve them and explicitly 
derive the hierarchy as well as solutions of the associated integrable non-linear ODEs provided 
that the form of the transform ${\mathbb M}$ is given. 

We consider below two types of Darboux transforms:

\begin{enumerate}

\item {\tt  Basic Darboux transform}

\begin{equation}
{\mathbb M}_n(\lambda) =  \lb \begin{matrix}
		\lambda {\mathbb I}^{({\cal N})} +A_n & B_n\\
		C_n &  \lambda {\mathbb I}^{({\cal M})} +D_n 
	\end{matrix} \rb = \lambda {\mathbb I} + {\cal K}_n. \label{typeII}
\end{equation}
The above will be used to provide the Darboux-BT relations and hence we can derive in 
straightforward manner the hierarchy as well as associated solutions.

\item {\tt NLS/Toda type Darboux}

\begin{equation}
{\mathbb M}_n(\lambda) =  \lb \begin{matrix}
		\lambda {\mathbb I}^{({\cal N})} +A_n & B_n\\
		C_n & \rho {\mathbb I}^{({\cal M})}
	\end{matrix} \rb, \label{typeI}
\end{equation}
where $\rho$ is a constant and for our proposes here will be set equal to zero (Toda-type Darboux).
\end{enumerate}
The entries of the matrices above are as follows: $B$ is an ${\cal N} \times {\cal M}$ matrix, 
$C$ is ${\cal M} \times {\cal N}$, $A$ is ${\cal N} \times {\cal N}$
and $D$ is ${\cal M} \times {\cal M}$.

\subsection{Conserved quantities}
\noindent 
We first focus on the derivation of the associated conserved quantities, which are generated by the transfer matrix defines as
\begin{equation}
{\mathfrak t}(\lambda) = tr \big (T(\lambda) \big ),~~~~T(\lambda) = L_N(\lambda) \ldots L_1(\lambda).\label{transfer}
\end{equation}
It is straightforward to show using also the discrete zero curvature condition: 
\begin{eqnarray}
\dot {\mathfrak t}(\lambda) =tr\big ( \sum_{n=1}^N L_{N} \ldots \dot L_n \ldots L_1\big ) = \ldots  = tr\big (V_{N+1}(\lambda)T(\lambda) - 
T(\lambda) V_1(\lambda) \big).
\end{eqnarray}
The latter expression is zero if we impose periodic boundary conditions $V_{N+1} = V_1$. 
Thus ${\mathfrak t}(\lambda) = \sum_{k=0}^N {I^{(k)} \over \lambda^k}$ is the generating function of the integrals of motion. 
In particular,  by expanding $\ln {\mathfrak t} = \sum_{k=0}^N {H^{(k)} \over \lambda^k}$  we find the local conserved quantities.
We keep here terms up to fourth order in the expansion of $ \ln{\mathfrak{t}}$ and obtain the first few local integrals of motion of the DNLS model:
\begin{eqnarray}
H_1 &=& tr\sum_{n=1}^N \mathbb{N}_n \label{eq:ALP_ExDNLSN} \nonumber\\
H_2 &=& tr \sum_{n=1}^N\Big (  X_nY_{n - 1} - \frac{1}{2}  \mathbb{N}_n^2 \Big ) \nonumber\\
H_3 &=&tr \sum_{n=1}^N  \Big ( X_n Y_{n - 2} -  \lb \mathbb{N}_n + \mathbb{N}_{n- 1} \rb X_n Y_{n - 1} + 
\frac{1}{3}  \mathbb{N}_n^3 \Big )\nonumber\\
H_4 &=& tr \sum_{n=1}^N \Big ( X_nY_{n-3} - \big ({\mathbb N}_{n-2} +{\mathbb N}_{n-1}+{\mathbb N}_{n}\big )
X_n Y_{n-2} +{\mathbb N}_{n-1}{\mathbb N}_{n}
X_n Y_{n-1} \nonumber\\
&  &+\  \big ({\mathbb N}^2_{n-1} +{\mathbb N}^2_{n}\big) X_nY_{n-1}- {1\over 2} X_n^2 Y^2_{n-1} - 
 X_n Y_{n-1} X_{n-1} Y_{n-2}-{1\over 4} {\mathbb N}_n^4\Big )\nonumber\\
& & \ldots
\label{eq:ALP_ExDNLSP} 
\end{eqnarray}
The general recursion relation for the charges in involution is presented in the appendix.

Recall $X_n,\ Y_n$ are matrices with entries $X_{n}^{ij},\ Y_n ^{ij}$ respectively. Requiring Poisson commutativity 
for the conserved quantities leads to a  $\mathfrak{gl}_{\cal N+ {\cal M}}$  generalization of 
the canonical commutation relations 
of the scalar DNLS model (see also \cite{DoFiSk1}):
\begin{equation}
\Big \{ X_{n}^{ij},\ Y_m^{kl}\Big \} = -\delta_{il} \delta_{jk} \delta_{nm}. \label{gl}
\end{equation}
The above Poisson structure is further verified when the equations of motion are computed via the Lax pair representation. 
Indeed, comparison between the
Hamiltonian approach  i.e. the equations of motions obtained  via $\dot \Phi = \{H,\ \Phi\}$ ($\Phi \in \{ X^{ij},\ Y^{ij}\}$) 
and the Lax pair description,  which will be discussed later in the text, leads to agreement, provided that (\ref{gl}) are considered.

\subsection{Dressing: Lax pairs and  solutions}
\noindent 
In this subsection we derive the ``dressed'' quantities  $V^{(\alpha)}$ of the 
hierarchy\footnote{We consider here for simplicity, but without really losing generality $\theta =1$ in ${\mathbb N} = \theta +XY$.}, 
and also identify solutions of the tower of the emerging non-linear ODEs.

\subsubsection{Basic Darboux}

\noindent We focus now on the basic Darboux matrix (\ref{typeII}) and on the time independent part of the transform 
(\ref{BT1}) and obtain stationary solutions (solitons, traveling waves); the time dependence is dictated by (\ref{BT2}).
From (\ref{BT1}) we obtain the following fundamental constraints:
\begin{eqnarray}
&& B_n = \hat X_n - X_n, ~~~~~C_{n+1} = Y_n - \hat Y_n\nonumber\\
&& B_{n+1} -B_ n = X_n Y_n B_n +X_n D_n -A_{n+1} \hat X_n \nonumber\\
&& C_{n+1} -C_n = -C_{n+1}\hat X_n \hat Y_n+ Y_n A_n -D_{n+1} \hat Y_n\nonumber\\
&& A_{n+1} -A_n = X_n Y_n -\hat X_n \hat Y_n ~~~~~D_{n+1} -D_n = -Y_n X_n +\hat Y_n \hat X_n. \label{basic0}
\end{eqnarray}
We moreover require that the Darboux matrix for $\lambda =0$ satisfies the generic 
quadratic relation, $M_n^2 = \hat \xi M_n + \zeta{\mathbb I}$, which leads to:
\begin{equation}
A_n^2 + B_n C_n = \hat \xi A_n + \zeta, ~~~~~~ D_n^2 +C_n B_n= \hat \xi D_n +\zeta. \label{constraint1}
\end{equation}

This is a fundamental case to consider, and although simple it fully describes the dressing process for the construction of the Lax pairs. 
From the basic relations (\ref{basic0}) we obtain  ($\hat X = \hat Y =0$):
\begin{eqnarray}
&& B_n = - X_n, ~~~~~C_{n+1} = Y_n ,  ~~~~B_{n+1} -B_ n = X_n Y_n B_n +X_n D_n, \nonumber\\
&& C_{n+1} -C_n = Y_n A_n, ~~~~A_{n+1} -A_n = X_n Y_n, ~~~~D_{n+1} -D_n = -Y_n X_n. \label{constr2b}
\end{eqnarray}
Note that time dependence for the fields is always implied.

We now focus on the dressing process in order to derive the time components of the Lax pairs.
Let $L^0,\ V^{(\alpha),0}$ be the ``bare'' Lax pairs:
\begin{equation}
L^0(\lambda) = {\mathbb I} + {\lambda \over 2}\big (\Sigma+ {\mathbb I}\big ),
~~~~V^{(\alpha), 0}(\lambda)= {\lambda^{\alpha} \over 2}\Sigma,
\end{equation}
where we define,  ${\mathbb I} = {\mathbb I}^{({\cal N} +{\cal M})}$ and
\begin{equation}
\Sigma =  \lb \begin{matrix}
		{\mathbb I}^{({\cal N}) } & 0 \\
		 0 & -{\mathbb I}^{({\cal M})}
	\end{matrix} \rb.
\end{equation}
Also, the ``dressed'' time components of the Lax pairs can be expressed as formal series expansions
\begin{equation}
V_n^{(\alpha)}(\lambda, t) = {\lambda^{\alpha} \over 2}\Sigma + \sum_{k=0}^{\alpha-1}\lambda^k w^{(\alpha)}_{n,k}(t),
\end{equation}
where the quantities $ w^{(\alpha)}_k$ will be identified via the dressing transform.
From the time part (\ref{BT2}) we obtain a set of recursion relations,  i.e.
\begin{eqnarray}
&& w^{(\alpha)}_{\alpha-1}= {1\over 2}\big [{\cal K},\ \Sigma\big ], \nonumber \\
&& w^{(\alpha)}_{k-1} =-w^{(\alpha)}_{k}\  {\cal K}, ~~~~k\in\{1, 2, \ldots, \alpha-1\}\nonumber\\
&& \partial_{t_{\alpha}} {\cal K} = w^{(\alpha)}_0\  {\cal K}. \label{recursion1}
\end{eqnarray}
We solve the latter recursion relations, and identify the first few time components of the Lax pairs:
 $V^{(\alpha)}$ operator is  identified  as $V^{(\alpha)} = \lambda V^{(\alpha-1)} + w_0^{(\alpha)}$. 
 Moreover, the recursion relations (\ref{recursion1})  lead to
\begin{equation}
w_k^{(\alpha)} = w_{k-1}^{(\alpha-1)}, ~~~~~w_{0}^{(\alpha)} = (-1)^{\alpha-1} w_{0}^{(1)}{\cal K}^{\alpha-1},
\end{equation}
where the latter relations together with the constraints (\ref{constr2b}) suffice to provide $w_0^{(\alpha)}$ at each order. 

We present below the first few members of the discrete hierarchy:

\begin{enumerate}

\item{\tt Discrete matrix DST-like equation}
\begin{eqnarray}
&& V_n^{(1)} =  \lb \begin{matrix}
		{\lambda \over 2} {\mathbb I}^{({\cal N})} & X_n\\
		Y_{n-1} &  - {\lambda\over 2} {\mathbb I}^{({\cal M})}
	\end{matrix} \rb.
\end{eqnarray}
The zero curvature conditions for $t_1$ leads to the equations of motion:
\begin{eqnarray}
&& \partial_{t_1 } X_n=  X_{n+1} -X_n -X_n Y_n X_n  \label{E11} \\
&&  \partial_{t_1 } Y_n= Y_{n} -Y_{n-1} +Y_n X_n Y_n. \label{E12}
\end{eqnarray}

\item{\tt Discrete matrix  NLS equation}

\begin{eqnarray}
 V_n^{(2)} =  \lb \begin{matrix}
		{\lambda^2 \over 2} {\mathbb I}^{({\cal N})} - X_n Y_{n-1} & \lambda X_n + X_{n+1} - {\mathbb N}_n X_n\\
		\lambda Y_{n-1} +Y_{n-2} - Y_{n-1} {\mathbb N}_{n-1} &  - {\lambda^2\over 2} {\mathbb I}^{({\cal M})}   +Y_{n-1}X_n
	\end{matrix} \rb,
\end{eqnarray}
and the zero curvature condition for $t_2$ leads to:
\begin{eqnarray}
	&& \partial_{t_2 } X_n = -X_{n + 1} Y_n X_n - \lb \mathbb{N}_n + \mathbb{N}_{n + 1} \rb  X_{n + 1}+ 
 \mathbb{N}_n^2X_n  - X_n Y_{n - 1}X_n + X_{n + 2} \nonumber\\ && \label{E21}\\
	&& \partial_{t_2 } Y_n =  Y_n  X_n Y_{n - 1}+ Y_{n- 1} \lb \mathbb{N}_n + \mathbb{N}_{n - 1} \rb-
 Y_n \mathbb{N}_n^2 +  Y_n X_{n + 1} Y_n- Y_{n- 2}. \nonumber\\ && \label{E22}
\end{eqnarray}

\item{\tt Discrete matrix generalized mKdV equation}
\begin{eqnarray}
&& V_n^{(3)} =  \lb \begin{matrix}
		{\lambda^3 \over 2} {\mathbb I}^{({\cal N})} -\lambda X_n Y_{n-1} & \lambda^2 X_n + \lambda X_{n+1} - \lambda {\mathbb N}_n X_n\\
		\lambda^2  Y_{n-1}+\lambda Y_{n-2} - \lambda Y_{n-1} {\mathbb N}_{n-1} &  - {\lambda^3\over 2} {\mathbb I}^{({\cal M})} +\lambda Y_{n-1}X_n 
	\end{matrix} \rb  +  w_{n,0}^{(3)}, \nonumber\\ && \label{fourth}
\end{eqnarray}
$w_{n,0}^{(3)}$ is  given by a rather lengthy expression in the appendix. 
\end{enumerate}
It is worth noting that the generalized zero curvature condition:
$\partial_{t_3} V_n^{(1)}- \partial_{t_1} V_n^{(3)}= \big [V_n^{(3)},\ V_n^{(1)} \big ]$,
provides the matrix discrete mKP-like equation. This significant  issue will be discussed in detail 
elsewhere together with the continuous case \cite{DoFiSk1}.

\subsection*{Solutions: 1-soliton}

\noindent In order to obtain solutions of the non-linear ODEs emerging from the DNLS hierarchy 
we focus on the case where $\hat X_n = \hat Y_n =0$. Then the basic relations (\ref{basic0}) reduce to
\begin{eqnarray}
&& \Delta X_n =  -X_n D_{n+1}, ~~~~~~ \Delta Y_{n-1} = Y_n A_n  \label{basic1}\\
&& \Delta D_n  =-Y_n X_n, ~~~~~~ \Delta A_{n} = X_n Y_n, \label{basic2}
\end{eqnarray}
where we define $\Delta F_n = F_{n+1} - F_n$. And the constraints (\ref{constraint1}) become
\begin{equation}
D_{n+1}^2 - Y_n X_{n+1} = \hat \xi D_{n+1} +\zeta, ~~~~~~~A_n^2 -X_n Y_{n-1} = \hat \xi A_n +\zeta. \label{constraint2}
\end{equation}
We distinguish two cases below, which provide two types of solitons:

\begin{enumerate}

\item Let us solve the equations involving $D_n,\ X_n$, we first solve the equation for $D_n$, and we consider $\zeta =0$,
\begin{eqnarray}
\Delta D_{n} =-Y_n X_n D_{n+1} - D_{n+1}^2 +\hat \xi D_{n+1} \ \Rightarrow\  D_{n} = \big ( \xi +D_n \big )D_{n+1}, \label{dd}
\end{eqnarray}
where $ \xi = 1 -\hat \xi$.
Before we proceed with the solution we consider also the following ansatz for the matrix fields:
\begin{equation}
X_n = {\mathrm x}_n \hat {\mathrm B}, ~~~~~Y_n ={\mathrm y}_n {\mathrm B}, ~~~~A_n = {\mathrm a}_n \hat {\mathrm B}{\mathrm B}, ~~~~~
D_n = {\mathrm d}_n  {\mathrm B} \hat {\mathrm B},  \label{ansatz1}
\end{equation}
where $\hat {\mathrm B}$ and ${\mathrm B}$ are ${\cal N}\times {\cal M}$ and ${\cal M} \times {\cal N}$ matrices respectively  that obey:
\begin{equation}
\hat {\mathrm B} {\mathrm B} \hat {\mathrm B} = \kappa \hat {\mathrm B}, ~~~~ 
{\mathrm B} \hat {\mathrm B}  {\mathrm B} = \kappa {\mathrm B}. \label{TL}
\end{equation}

Indeed, from (\ref{dd}) and (\ref{ansatz1}), (\ref{TL}) we conclude:
\begin{equation}
{\mathrm d}_{n+1} = {{\mathrm d}_n \over  \xi + \kappa {\mathrm d}_n}\  \Rightarrow\ {\mathrm d}_{n} = 
{( \xi -1){\mathrm d}_1 \over \xi^{n-1} (\xi -1)+(\xi^{n-1} -1) \kappa {\mathrm d}_1}, \label{dd2}
\end{equation}
and periodic conditions are ensured provided that $\xi^N  =1$. Then from (\ref{basic1}) and (\ref{ansatz1}) 
and knowing the solutions for ${\mathrm d}_n $ (\ref{dd2})  we conclude
\begin{eqnarray}
{\mathrm x}_{n} =  \prod_{m=2}^n (1-\kappa {\mathrm d}_{m}){\mathrm x}_1\ \Rightarrow\  
{\mathrm x}_n = {\xi^{n-1} ( \xi -1){\mathrm x}_1 \over \xi^{n-1}(\xi -1+ {\kappa }{\mathrm d}_1) -{\kappa }{\mathrm d}_1 }. \label{x1}
\end{eqnarray}

Similarly,  for $Y_n, \ A_n$ we first identify the solution for $A_n$ and then we derive the one soliton solution for $Y_n$.
Focus on relations (\ref{basic1})--(\ref{constraint2}) for the pair $A_n,\  Y_n $:
\begin{equation}
{\mathrm a}_{n+1} -{\mathrm a}_n = \kappa {\mathrm a}_{n+1} {\mathrm a}_n - \hat \xi {\mathrm a}_n\  \Rightarrow\  
{\mathrm a}_{n+1} = {{\mathrm a}_{n} \over \tilde \kappa {\mathrm a}_{n} + \tilde \xi}, \label{rec2}
\end{equation}
where $\tilde \xi= \xi^{-1}$ and $\tilde \kappa= -\kappa \xi^{-1}$.
The latter leads to the following solution:
\begin{equation}
{\mathrm a}_{n} = {( \xi -1 ){\mathrm a}_1 \over  \xi^{-n+1}(\xi -1) +  \kappa {\mathrm a}_1( \xi^{-n+1} -1)}. \label{asol}
\end{equation}
From (\ref{basic1}), (\ref{asol}) we conclude
\begin{eqnarray}
&& {\mathrm y}_{n} = \prod_{m=2}^n \big (1 -\kappa {\mathrm a}_m \big )^{-1} {\mathrm y}_1\  \Rightarrow \ {\mathrm y}_n = 
{  \xi^{-n}(\xi -1)(1- \kappa {\mathrm a}_1) {\mathrm y}_1 \over  \xi^{-n}(\xi -1 + \kappa {\mathrm a}_1)  -\kappa {\mathrm a}_1}. \label{y1}
\end{eqnarray}
Periodic  boundary conditions are valid for all the associated fields and this can be easily checked by inspection provided that $ \xi^{N} =1$. Note that
 in the stationary solutions above the time dependence, for each time flow,  is naturally introduced:  $ \xi^{n} \to  \xi^{n} e^{\Lambda^{(\alpha)} t_{\alpha}}$, where 
$\Lambda^{(\alpha)} = C_{\alpha}(\xi -1)^{\alpha}$, (see also next subsection and last section where a detailed discussion on the related dispersion relations is presented).

\item We now consider $\hat \xi =0$ and $\zeta \neq 0$ in (\ref{constraint2}). 
We define $\hat D_n = D_n+ c{\mathbb I}^{(\cal M)},\  \hat A_n = A_n + c {\mathbb I}^{(\cal N)}$,
where $\zeta =c^2$.
We consider the ansatz (\ref{ansatz1}) for $X_n$ and $Y_n$, and also
\begin{equation}
\hat A_n =\hat  {\mathrm a}_n \hat {\mathrm B}{\mathrm B}, ~~~~~ 
\hat D_n = \hat {\mathrm d}_n  {\mathrm B} \hat {\mathrm B}. \label{ansatz21}
\end{equation}
The matrices ${\mathrm B},\ \hat {\mathrm B}$ satisfy: $\hat {\mathrm B} {\mathrm B} = 
\kappa {\mathbb I}^{(\cal N)},\  {\mathrm B} \hat {\mathrm B} = \kappa {\mathbb I}^{(\cal M)}$, and we also set 
$D_n = {\mathrm d}_n {\mathbb I}^{(\cal M)},\ A_n = {\mathrm a}_n {\mathbb I}^{(\cal N)}$.
We then conclude (\ref{basic1})--(\ref{constraint2}), $\hat \xi =0,\ \zeta\neq 0$:
\begin{equation}
\hat {\mathrm d}_{n+1}= {\hat {\mathrm d}_n \over \bar \kappa \hat {\mathrm d}_n + \bar \xi}, ~~~~~
\hat {\mathrm a}_{n+1}= {\hat {\mathrm a}_n \over \tilde \kappa \hat {\mathrm a}_n + \tilde \xi} \label{ad2}
\end{equation}
where $\bar \xi = \epsilon \eta^{-1},\ \bar \kappa = \kappa \eta^{-1}$, $~\eta = 1 +c,\ \epsilon = 1-c$, 
and $\tilde \xi = \bar \xi^{-1},\ \tilde \kappa = -\bar \kappa \bar \xi^{-1}$, and clearly 
$\hat {\mathrm d}_n = \kappa^{-1}({\mathrm d}_n +c),\ \hat {\mathrm  a}_n=\kappa^{-1}({\mathrm a}_n +c),$
are given by  (\ref{dd2}), (\ref{asol}), but with $\xi \to \bar \xi,\ \kappa \to \bar \kappa$.
From the general expressions (\ref{dd2}), (\ref{x1}), (\ref{rec2})-(\ref{y1}):
\begin{eqnarray}
&& {\mathrm x}_n = {(\bar \xi -1) {\mathrm x}_1 \over (\bar \xi -1 +\bar \kappa \hat {\mathrm d}_1) \eta ^{-n+1}  - \bar \kappa \hat {\mathrm d}_1
\epsilon^{-n+1} }, \nonumber\\
&& {\mathrm y}_n = {\eta (\tilde \xi -1)(1-\tilde \kappa \hat {\mathrm a}_1 ) {\mathrm y}_1 \over (\bar \xi - 1 +\bar \kappa \hat {\mathrm a}_1) \eta ^{n}  - 
\bar \kappa \hat  {\mathrm a}_1\epsilon^{n}}.\label{xy2}
\end{eqnarray}
As in the previous case the time dependence is easily implemented: $\eta^{n}\to \eta^{n}e^{ \Lambda^{(\alpha)}t_{\alpha}},\ \epsilon^{n}\to 
\epsilon^{n}e^{\hat \Lambda^{(\alpha)}t_{\alpha}}$, $\hat \Lambda^{(\alpha)}=\hat c_{\alpha} (\epsilon -1)^{\alpha} ,\ \Lambda^{(\alpha)}= 
c_{\alpha}(\eta-1)^{\alpha}$ (see also next subsection).
\end{enumerate}

\subsubsection{Toda type Darboux}

\noindent We now present the Darboux-BT relations associated to the Toda type transform (\ref{typeI}). 
After solving the set of equations provided by (\ref{BT1}) for (\ref{typeI}) we conclude
\begin{eqnarray}
&& B_n =\hat X_n, ~~~~ C_{n+1} = Y_n,~~~~~A_{n+1} \hat X_n +B_{n+1} = {\mathbb N}_n B_n \nonumber\\
&& C_{n+1} \hat {\mathbb N}_n = Y_n A_n +C_n, ~~~~~A_{n+1} -A_n = {\mathbb N}_n - \hat {\mathbb N}_n. \label{setI}
\end{eqnarray}
To obtain solutions we focus  on the special case where $\hat Y_n =0$, we then observe that $\hat X_n$ automatically
satisfies the linearizations of the non-linear integrable ODEs (\ref{E11}), (\ref{E21}) and so on (for $Y_n=0$).  
We also consider the ansatz (\ref{ansatz1}), (\ref{TL}) 
for the fields as well as: $\hat X_n = \hat {\mathrm x}_n \hat {\mathrm B}$,
where $\hat {\mathrm x}_n$ then satisfies
\begin{eqnarray}
&& \partial_{t_1} \hat {\mathrm x}_n = \hat {\mathrm x}_{n+1} - \hat {\mathrm x}_n, \label{lin1} \\
&& \partial_{t_2} \hat {\mathrm x}_n = \hat {\mathrm x}_{n+2} -2 \hat {\mathrm x}_{n+1} + \hat {\mathrm x}_n, \label{dheat} \\
&& \ldots \nonumber\\
&& \partial_{t_{\alpha}} \hat {\mathrm x}_n  = \sum_{k=0}^{\alpha} (-1)^{\alpha-k} \lb \begin{matrix}
		\alpha \\
		k
	\end{matrix} \rb\hat {\mathrm x}_{n+k}. \label{discr1}
\end{eqnarray}
The first of the equations above is the discrete transport equation, the second one is the discrete heat equation and so on.
The solutions of the linear equations have the generic form
\begin{equation}
\hat {\mathrm x}_n =  \sum_{s=1}^S c_s \xi_s^{n-1} e^{ \sum_{\alpha}\Lambda_s^{(\alpha)} t_{\alpha}}, ~~~\mbox{and/or} ~~~
\hat {\mathrm x}_n  = \int_{|\xi|=1} d\xi \  c(\xi) \xi^{n-1} e^{\sum_{\alpha} \Lambda_{\xi}^{(\alpha)} t_{\alpha}}, \label{four2}
\end{equation}
and the associated dispersion relations are easily extracted in this setting and read as
\begin{equation}
\Lambda_s^{(\alpha)} =( \xi_s -1)^{\alpha}.\label{disp}
\end{equation}

The set of equations (\ref{setI}) reduce to
\begin{eqnarray}
&& {\mathrm y}_n -{\mathrm y}_{n-1} = \kappa {\mathrm y}_n {\mathrm a}_n, \label{1} \\
&& \hat  {\mathrm x}_{n+1} \hat {\mathrm x}_n^{-1} -1 =\kappa ({\mathrm x}_n {\mathrm y}_n - {\mathrm a}_{n+1} )\label{2}\\
&& {\mathrm a}_{n+1} -{\mathrm a}_n = {\mathrm x}_n {\mathrm y}_n. \label{3}
\end{eqnarray}
Via (\ref{1})--(\ref{3}) we obtain 
\begin{eqnarray}
{\mathrm y}_ n = \prod_{m=2}^n \big ( 1 - \kappa {\mathrm a}_m \big )^{-1} {\mathrm y}_1 , ~~~~
{\mathrm x}_ n = ({\mathrm a}_{n+1} -{\mathrm a}_n)  \prod_{m=2}^n \big ( 1 - \kappa {\mathrm a}_m \big ) {\mathrm y}_1^{-1}.
\end{eqnarray}
Having the solution $\hat {\mathrm x}_n$ at our disposal we can immediately solve for 
\begin{equation}
1- \kappa {\mathrm a}_n =\hat {\mathrm x}_{n+1} \hat {\mathrm x}_{n}^{-1}\ \Rightarrow\ 
\prod_{m=2}^n \big ( 1 -  \kappa{\mathrm a}_m \big ) = {\hat {\mathrm x}_{n+1}\over \hat {\mathrm x}_2}, \label{aa}
\end{equation}
and hence obtain the explicit expressions for both fields:
\begin{eqnarray}
{\mathrm y}_ n ={\hat {\mathrm x}_2 \over \hat {\mathrm x}_{n+1}} {\mathrm y}_1, ~~~~~~
{\mathrm x}_ n = -\kappa^{-1} \hat {\mathrm x}_2^{-1} {\mathrm y}_1^{-1}
{\hat {\mathrm x}_{n+2} \hat {\mathrm x}_n - \hat {\mathrm x}_{n+1}^2 \over \hat {\mathrm x}_{n}}.\label{y}
\end{eqnarray}
Periodic boundary conditions (${\mathrm x}_{N+1} = {\mathrm x}_1,\ {\mathrm y}_{N+1} = {\mathrm y}_1$) are valid provided that 
$\hat {\mathrm x}_{N+1} =\hat {\mathrm x}_1$.
Also, $\hat {\mathrm x}_2$ (boundary term) in the expressions above is treated as a constant i.e. $\partial_t \hat {\mathrm x}_2 =0$.
Expressions  (\ref{y}) are general new solutions
of the non-linear ODEs for the DNLS hierarchy in terms of solutions of the associated linear equations.
We describe below two simple solutions of the type (\ref{y}), which reproduce the two types 
of solitons discussed in the previous subsection.

\begin{itemize}
\item {\tt Simple solutions}

We consider the following simple linear solutions:
\begin{enumerate}
\item We first choose:
\begin{eqnarray}
&& \hat {\mathrm x}_n = c_1 + c_2 \xi^{n-1} e^{\Lambda^{(\alpha)}t_{\alpha}} \label{11}\\
&& \Lambda^{(\alpha)} = (\xi-1)^{\alpha} \label{disp1}
\end{eqnarray}
see also (\ref{disp}). We substitute (\ref{11})  in (\ref{y}), 
and we recover solution (\ref{x1}), (\ref{y1}):
\begin{eqnarray}
{\mathrm x}_n  = -{\kappa^{-1}\hat {\mathrm x}_2^{-1} {\mathrm y}_1^{-1}c_1 c_2 (\xi-1)^2\over c_2 + 
c_1 \xi^{-n+1} e^{-\Lambda^{(\alpha)} t_{\alpha}}}, ~~~~~  {\mathrm y}_n  = {\hat {\mathrm x}_2 {\mathrm y}_1\over c_1+ 
c_2 \xi^{n} e^{\Lambda^{(\alpha)} t_{\alpha}}}. \label{soliton1}
\end{eqnarray}
\item The second simple choice is:
 \begin{eqnarray}
&& \hat {\mathrm x}_n = c_1\eta^{n-1} e^{\Lambda^{(\alpha)}t_{\alpha}} +  c_2\epsilon^{n-1} e^{\hat \Lambda^{(\alpha)}t_{\alpha}}  \\
&&  \Lambda^{(\alpha)} = (\eta -1)^{\alpha},\ ~~~~~\hat\Lambda^{(\alpha)} = (\epsilon -1)^{\alpha}, \label{disp2}
\end{eqnarray}
see also (\ref{disp}).
After substituting the above in (\ref{y}) we recover (\ref{xy2}):
\begin{eqnarray}
&& {\mathrm x}_n  = -{\kappa^{-1}\hat {\mathrm x}_2^{-1} {\mathrm y}_1^{-1}c_1 c_2 (\eta - \epsilon)^2\over 
c_1 \epsilon^{-n+1} e^{-\hat \Lambda^{(\alpha)} t_{\alpha}} + c_2\eta^{-n+1}e^{-\Lambda^{(\alpha)}t_{\alpha}}} 
\nonumber\\
&&  {\mathrm y}_n  = {\hat {\mathrm x}_2 {\mathrm y}_1 \over  c_1 \eta^{n} e^{\Lambda^{(\alpha)} t_{\alpha} } +
c_2 \epsilon^{n} e^{\hat \Lambda^{(\alpha)} t_{\alpha}  }}. \label{soliton2}
\end{eqnarray}
\end{enumerate}
Notice that we easily  recover here the solutions of the previous subsection  
with a different, rather more convenient parametrization. We have not assumed any extra constraints 
for the fields contrary to the previous case,
it is thus clear that the computation in this case is more straightforward, provided that the 
form of Fourier transform is available. 
The Fourier coefficients, and in turn the solutions of the non-linear ODEs, 
can be expressed in terms of some given initial profile, and  different choices of initial 
profiles will provide distinct general solutions (for relevant discussion in the continuous case  
see the following Remark and
also \cite{Ablo2, ZakharovShabat, BDMS} and references therein). 
This is a very interesting problem, especially in the discrete case, and will be fully
investigated elsewhere.
\end{itemize}

\noindent {\bf Remark 1: the continuous case.}\\
Let us briefly discuss the continuous case using the Toda type Darboux transformation (\ref{typeI}). 
Recall the $U$-operator (see also \cite{DoFiSk1}) of the continuous Lax pair $(U,\ V)$
\begin{equation}
U(x, \lambda )=  \lb \begin{matrix}
		{\lambda \over 2}{\mathbb I}^{({\cal N})} & \hat {\cal V}\\
		{\cal V }&  - {\lambda\over 2} {\mathbb I}^{({\cal M})}
	\end{matrix} \rb \label{uu}
\end{equation}
We also consider the ansatz:  $\hat{\cal  V} = \hat u \hat {\mathrm B}$,  
${\cal V} = u {\mathrm B}$ and as in the discrete case ${\mathrm B},\ \hat {\mathrm B}$ satisfy (\ref{TL}). 
The Darboux matrix is the Toda type matrix (\ref{typeI}) parametrized now by the continuous $x$-variable 
(time dependence is always implicit): 
\begin{equation}
{\mathbb M}(x,  \lambda) =  \lb \begin{matrix}
		\lambda {\mathbb I}^{({\cal N})} +A(x) & B(x)\\
		C(x) & \rho {\mathbb I}^{({\cal M})}
	\end{matrix} \rb.\label{typeIb}
\end{equation}
The $x$-part of the Darboux transform (see e.g \cite{DoFiSk1} and references therein) gives:
\begin{equation}
\partial_x {\mathbb M}(x, \lambda) =  U(x, \lambda) {\mathbb M}(x, \lambda)- {\mathbb M}(x, \lambda) U_0(x, \lambda), \label{xxdarboux}
\end{equation}
where $U_0$ is  also given by (\ref{uu}), but ${\cal V} \to {\cal V}_0,\ \hat {\cal V} \to \hat {\cal V}_0$.
If ${\cal V}_0 =0 $, and also $\hat {\cal V}_ 0 =  \hat u_0 \hat {\mathrm B}$,
then $\hat u_0$ satisfies the linear equations (we will consider below the example of the NLS-like equations. 
For details on the $V$-operators for the hierarchy we refer the interested reader in \cite{DoFiSk1}):
\begin{equation}
\partial_{t_\alpha} \hat u_0 =\partial_x^{\alpha} \hat u_0, \label{linearn}
\end{equation}
for each time flow $t_{\alpha}$. Consequently, $\hat u_0$ can be expressed as
\begin{equation}
\hat u_0 = \sum_{s=1}^S c_s e^{-k_s x + \sum_{\alpha} \Lambda_s^{(\alpha)}t_{\alpha}}, ~~~~\mbox{and/or} ~~~~~
\hat u_0 = \int_{\mathbb R} d\lambda c(\lambda) e^{i\lambda x + \sum_{\alpha} \Lambda_{\lambda}^{(\alpha)}t_{\alpha} }\label{fourier2}
\end{equation}
with dispersion relations given as
\begin{equation}
\Lambda^{(\alpha)}_s = (-k_s)^{\alpha}, ~~~~~\Lambda^{(\alpha)}_{\lambda} = (i \lambda)^{\alpha}.
\end{equation}

From the Darboux relations (\ref{xxdarboux}), and  assuming $A = a \hat {\mathrm B} {\mathrm B}$, we obtain:
\begin{eqnarray}
\partial_x \hat u_0 = - \kappa a \hat u_0, ~~~~\partial_x u =  \kappa a u, 
~~~~~\partial_x a =   u \hat u. \label{basis}
\end{eqnarray}
Solving the equations above leads to:
\begin{equation}
u= {g \over \hat u_0} ~~~~\mbox{and} ~~~~~\hat u = - \kappa^{-1}g^{-1}{\hat u_0 \partial_x^2(\hat u_0) - 
(\partial_x \hat u_0)^2 \over \hat u_0}, \label{contgen}
\end{equation}
which are the continuous analogues of (\ref{y}). Choosing for instance the simple linear solutions: 
$\hat u_0 = c_1 + c_2 e^{- kx + \Lambda^{(\alpha)}t_{\alpha}}$  
or  $\hat u_0 = c_1 e^{- k_1x + \Lambda_1^{(\alpha)}t_{\alpha}}+ c_2 e^{- k_2x + \Lambda_2^{(\alpha)}t_{\alpha}}$ 
we obtain one soliton solutions,  i.e. the continuous analogues of (\ref{soliton1}), (\ref{soliton2}).

Let us also consider examples of more general solutions. For instance in the NLS case the fields satisfy:
\begin{eqnarray}
&& \partial_t u+\partial_x^2 u -2\kappa \hat u u^2=0, \label{nlslike}
\end{eqnarray}
and similarly, for $\hat u$ with $t \to -t$.
$\hat u_0$  is a solution of the heat equation, $\alpha=2$ in (\ref{linearn}).
The general solution of the heat equation is expressed as  in (\ref{fourier2}) with dispersion relation $\Lambda_{\lambda}^{(2)} =- \lambda^2$.
Let $f_{0}$ be any initial profile for the linear solution, then $c(\lambda) = {1\over 2\pi}\int_{{\mathbb R}} d\xi e^{-i\lambda \xi}f_0(\xi)$,  
substituting $c(\lambda)$ in (\ref{fourier2}) and performing the Gaussian integral, 
due to the quadratic dispersion relation, we obtain:
\begin{equation}
\hat u_0 = {1 \over \sqrt{4 \pi t}}\int_{\mathbb R} d\xi e^{- {(x-\xi)^2 \over 4t}}f_0(\xi). \label{heat}
\end{equation}
Let us for instance consider the simple, although singular, initial profile $f_0(\xi) = \delta(\xi)$, 
then from (\ref{basis}) and (\ref{heat})  it follows that
\begin{equation}
u = gt^{1\over 2} e^{{x^2 \over 4t}} ~~~~ \mbox{and} ~~~~\hat u = \kappa^{-1} g^{-1} {e^{-{x^2 \over 4t}} \over 2 t^{3\over 2}}, \label{solh}
\end{equation}
where $g$ is an integration constant.
Substituting $u,\ \hat u$ above in (\ref{nlslike}) confirms that these are  indeed solutions of the NLS-type equations (\ref{nlslike}).
However, note that for the solution (\ref{solh}), $u \to \infty$ as $x \to \infty$, and also the behavior at $t=0$ is singular.
Different choices of initial profiles of the linear solution will naturally provide distinct solutions of the non-linear PDEs.

As in the discrete case this is a significant result, that allows the derivation of generic solutions of the non-linear PDEs 
in terms of linear solutions in a straightforward manner. 
Similarly,  due to the cubic dispersion relation for the generalized mKdV equation solutions of the linear problem can
be expressed in terms of Airy functions (see also e.g \cite{Ablo2, DoFiSk1}). $\square$

\noindent {\bf Remark 2: discrete Burgers equation.}\\
The discrete version of the viscous Burgers equation can be obtained from the discrete heat equation through 
the analogue of the Cole-Hopf transformation. Indeed,  we set 
$\hat {\mathrm x}_n = e^{{\mathrm y}_n}$,  in the discrete heat equation (\ref{heat}), then:
\begin{equation}
\partial_t {\mathrm y}_n = 
e^{\Delta{\mathrm y}_n}\Big (  e^{\Delta{\mathrm y}_{n+1}} -1\Big ) -\Big ( e^{\Delta{\mathrm y}_{n}}-1\Big ),  \label{HJ}
\end{equation}
where we define $\Delta f_n = f_{n+1} - f_{n}$,  $~\Delta^2 f_n = f_{n+2} - 2f_{n+1} + f_n$.
We also set $u_n = \Delta {\mathrm y}_{n}$, and we obtain a discrete version of the Burgers equation
\begin{equation}
\partial_t u_n =   e^{u_{n+1}} \Big ( e^{u_{n+2}}-  e^{u_{n}} \Big ) -2  \Big ( e^{u_{n+1}}-  e^{u_{n}} \Big ). \label{Burgers}
\end{equation}

Assuming the scaling $\Delta {y_n} \sim \delta$, we expand the exponentials and keep up to second order terms in (\ref{HJ}), (\ref{Burgers}):
\begin{eqnarray}
&& \partial_t{\mathrm y}_n = \Delta^{2} {\mathrm y}_n + \big (\Delta {\mathrm y}_n \big)^2 + {\cal O}(\delta^3 )\\
&& \partial_t u_n = \Delta^2 u_n + \Delta u_n^2 + {\cal O}(\delta^3).
\end{eqnarray}
The second of the equations above provides a good approximation for the discrete viscous  Burgers equation in the thermodynamic limit. $\square$

\subsection{Multiparticle-like solutions}
\noindent  The aim now is to identify more general solutions e.g. multi-solitons or bound states of two solitons i.e. breathers.
This can be achieved through various paths,  but we focus here on two distinct scenarios:

\begin{enumerate}
\item {\tt The general Darboux transform}

The fundamental Darboux transforms introduced above provide easily the one soliton solution,
however to obtain more general solutions of the non-linear ODEs above we  
implement  the general Darboux expressed as a formal 
$\lambda$-series expansion, in the typical loop group expansion fashion \cite{loop1},
\begin{equation}
{\mathbb M}_n(\lambda)= \lambda^m {\mathbb I}+ \sum_{k=0}^{m-1} \lambda^k g_n^{(k)}, \label{generalg1}
\end{equation}
where $g_{k}$ are $({\cal N}  +{\cal M})\times ({\cal N} +{\cal M})$ matrices to be identified via the Darboux transformation relations (\ref{BT1}).
We focus on the fundamental relations arising from the discrete space part of the Darboux transform (\ref{BT1}), 
where $\hat L_n = \lambda \Sigma^+ + {\mathbb I}$,  $~L_n = \lambda \Sigma^+ + {\mathbb I}+ U_n$ (\ref{DNLS}),  and we define
\begin{equation}
\Sigma^+ =  \lb \begin{matrix}
		 {\mathbb I}^{({\cal N})} & 0\\
		0 &  0
	\end{matrix} \rb, ~~~~~~ U_n =  \lb \begin{matrix}
		 X_n Y_n & X_n\\
		Y_n &  0
	\end{matrix} \rb.
\end{equation}
Then the following set of recursion relations emerge
\begin{eqnarray}
&& g_{n+1}^{(m-1)}\Sigma^+  -\Sigma^+  g_n^{(m-1)} = U_n \label{c1}\\
&& g_{n+1}^{(k)}  -g_n^{(k)} = \Sigma^+  g_n^{(k-1)} - g_{n+1}^{(k-1)}\Sigma^+ + U_ng_n^{(k)} \label{c2}\\
&& g_{n+1}^{(0)} -g_n^{(0)} = U_n g_n^{(0)}.
\end{eqnarray}
Specifically, the $m=2$ case provides for instance the two soliton solutions or the breather solution.
Solving the above recursion relations provides all $g_k$ as well as solutions of the associated non-linear ODEs. 
Extra constraints are required in order to be able to solve explicitly 
the algebraic relations  involved. This is particularly demanding computationally, therefore we focus below on a different, 
more practical methodology to obtain for instance the two-soliton solution.

\item {\tt Bianchi's permutability and the soliton lattice}

It is practically convenient to derive the two soliton or breather solution via Bianchi's permutatbility theorem, 
i.e. from the commutativity of Darboux transforms. Let us consider the following two families of Darboux transforms: 
\begin{enumerate}
\item ${\mathbb M}_n^{(i0)},\ i \in \{ 1,\ 2 \}$, which provides the one soliton solution: $F_n^{(i)},\  (F_n \in \{ X_n,\ Y_n\}) $, 
with parameter $\xi_i$ (see (\ref{constraint2}) $\zeta =0$)
 when considering  trivial initial solutions $F_n^{(0)} =0$.

\item  ${\mathbb M}_n^{(ij)}\ i, j \in \{ 1,\ 2 \}$, which provides
 the two-soliton solution: $F_n^{(ij)}$ with parameters $\xi_{i},\ \xi_j$  (\ref{constraint2}), when considering the one-soliton 
as initial solution: $F_n^{(j)}$, with parameter $\xi_j$.
\end{enumerate}

Commutativity of the  Darboux transforms requires:
$F_n^{(ij)} = F_n^{(ji)} = F_n$, more precisely:
\begin{equation}
{\mathbb M}^{(21)}_n(\lambda,\ \xi_2 ) \ {\mathbb M}^{(10)}_n(\lambda,\ \xi_1) = {\mathbb M}^{(12)}_n(\lambda,\ \xi_1 ) \ {\mathbb M}^{(20)}_n(\lambda,\ \xi_2),  \label{Bian}
\end{equation}
where we consider the fundamental Darboux transform (\ref{typeII}), i.e.
\begin{equation}
{\mathbb M}_n^{(kl)}(\lambda,\ \xi_k) = \lambda + K_n^{(kl)}(\xi_k), ~~~~~ K_n^{(kl)}(\xi_k) =  \lb \begin{matrix}
		A_n^{(kl)} & B_n^{(kl)}\\
		C_n^{(kl)} &  D_n^{(kl)}
	\end{matrix} \rb.
\end{equation}
Via the fundamental relations (\ref{Bian}) we then obtain:
\begin{eqnarray}
&& K_n^{(12)}= K_n^{(21)}  +K_n^{(1)}- K_n^{(2)}\nonumber\\
&& K_n^{(21)}K_n^{(1)}= K_n^{(12)}K_n^{(2)}\  \Rightarrow\  K_n^{(21)}\  \hat \Delta K_n\ = \hat \Delta K_n\ K_n^{(2)}, \label{factor0}
\end{eqnarray}
where we define in general, $\hat \Delta f_n = f_n^{(1)}- f_n^{(2)}$.

Now it is straightforward to solve the above algebraic relations and obtain the associated fields, i.e. the two-soliton solution. 
Indeed, from (\ref{factor0}) and recalling (\ref{basic0}) as well as the form of all solutions (\ref{ansatz1}) we obtain 
explicit analytic expressions for the two-soliton solution:
\begin{eqnarray}
&& {\mathrm x}_n = {\mathrm x}_n^{(1)} + {\kappa {\mathrm x}_n^{(2)} 
(\hat \Delta {\mathrm a}_n )^2+ {\mathrm y}_{n-1}^{(2)}(\hat \Delta {\mathrm x}_n)^2 -\kappa({\mathrm a}_n^{(2)} - 
{\mathrm d}_n^{(2)})\hat \Delta {\mathrm a}_n \hat \Delta {\mathrm x}_n \over 
\hat \Delta {\mathrm x}_n\hat \Delta {\mathrm y}_{n-1} +\kappa \hat \Delta {\mathrm a}_n \hat \Delta{\mathrm d}_n} \nonumber\\
&& {\mathrm y}_{n-1} ={\mathrm y}_{n-1}^{(1)} + {\kappa {\mathrm y}_{n-1}^{(2)} 
(\hat \Delta {\mathrm d}_n )^2+ {\mathrm x}_{n}^{(2)}(\hat \Delta {\mathrm y}_{n-1})^2 +\kappa({\mathrm a}_n^{(2)} - 
{\mathrm d}_n^{(2)})\hat \Delta {\mathrm d}_n \hat \Delta {\mathrm y}_{n-1} \over 
\hat \Delta {\mathrm x}_n\hat \Delta {\mathrm y}_{n-1} +\kappa \hat \Delta {\mathrm a}_n \hat \Delta{\mathrm d}_n}, \nonumber\\&& \label{2sol}
\end{eqnarray}
where the one soliton solutions ${\mathrm x}_n^{(i)},\ {\mathrm y}_n^{(i)}$ are given in (\ref{x1}), (\ref{y1}) 
and ${\mathrm d}_n, {\mathrm a}_n$ are given in (\ref{dd2}), (\ref{asol})  with corresponding parameters $\xi_i$. 
The second type of solitons are given in  (\ref{xy2}) and 
${\mathrm d}_n =\kappa \hat {\mathrm d}_n -c,\ {\mathrm a}_n =\kappa \hat {\mathrm a}_n -c $, with parameters $\eta_i,\ \epsilon_i$, 
see also (\ref{ad2}) and definitions below.

The same Darboux transform ${\mathbb M}_n \to {\mathbb M}(x)$ was also employed in the continuum case \cite{DoFiSk1}, 
thus the discussion above, and in particular relations (\ref{factor0}), (\ref{2sol})
are valid in the continuous case as well, given that:
\begin{equation}
{\mathrm x}_n \to \hat u(x),~~~~{\mathrm y}_n \to u(x),  ~~~~{\mathrm a}_n \to {\mathrm a}(x), ~~~~
{\mathrm d}_n \to {\mathrm d}(x) \label{Dict}
\end{equation}
where $\hat u,\ u$ are the matrix AKNS fields (following the notation of \cite{DoFiSk1}). 
The two soliton solutions are given by (\ref{2sol}) in terms of one soliton solutions $\hat u^{(i)}(x),\ 
u^{(i)}(x)$ (${\mathrm a}^{(i)}(x),\ {\mathrm d}^{(i)}(x)$) reported in
\cite{DoFiSk1}, subject to the ``dictionary'' (\ref{Dict}).
\end{enumerate}

\section{The matrix AL model}

\noindent  We come now to the study of the matrix AL model. As in the analysis presented in the previous section 
the main input is the $L$-operator,  which for the matrix AL model is of the form
\begin{equation}
L_n(z) = \lb \begin{matrix}
		z {\mathbb I}^{({\cal N})}& \hat b_n\\
		b_n &  z^{-1}{\mathbb I}^{({\cal M})} 
	\end{matrix} \rb, \label{AL}
\end{equation}
where $z$ is the multiplicative spectral parameter and $\hat b,\ b$ are ${\cal N} \times {\cal M}$ and ${\cal M} \times {\cal N}$ matrices respectively.
The conserved quantities can be obtained by expanding the monodromy matrix at $z\to \infty $ or $z \to 0$. 
By adding the first non trivial contributions  in the $z$ and $z^{-1}$ powers series expansion we get the associated Hamiltonian
as a linear combination of the $z$ and $z^{-1}$ expansion
\begin{equation}
H= H^+ + c  H^- =  \sum_{n=1}^N \big ( \hat b_{n+1} b_n + c b_{n+1} \hat b_n\big ), \label{HAL}
\end{equation}
$c$ is an arbitrary constant, and will be considered henceforth to be one.

We consider here, as in the previous section, two types of Darboux transforms (see aslo \cite{Suris}--\cite{Rourke}):

\begin{itemize}
\item {\tt The fundamental Darboux}

To construct the Lax pair as well as solitonic solutions we consider the fundamental Darboux matrix:
\begin{equation}
{\mathbb M}_n(z) = \lb \begin{matrix}
		Qz {\mathbb I}^{({\cal N})} - Q^{-1}z^{-1} A_n& B_n\\
		C_n &  QzD_n- Q^{-1} z^{-1}{\mathbb I}^{({\cal M})}
	\end{matrix} \rb \label{DAL1}
\end{equation}
where $Q = e^{\Theta}$ is an arbitrary constant, $A,\  D$ are 
${\cal N} \times {\cal N}$ and ${\cal M} \times {\cal  M}$ matrices respectively, and $B,\  C$ are 
${\cal N} \times {\cal M}$ and ${\cal M} \times {\cal N}$ matrices respectively.

\item {\tt The deformed oscillator-type  Darboux}

A different Darboux that has been more widely used both at classical and quantum level 
\cite{Suris, Korff, DoikouFindlay1} is given by
\begin{equation}
{\mathbb M}_n(z) = \lb \begin{matrix}
		Qz {\mathbb I}^{({\cal N})} - Q^{-1}z^{-1} A_n& B_n\\
		C_n & - Q^{-1} z^{-1}{\mathbb I}^{({\cal M})}
	\end{matrix} \rb, \label{DAL}
\end{equation}
$A$ is an ${\cal N} \times {\cal N}$  matrix, and $B,\  C$ are 
${\cal N} \times {\cal M}$ and ${\cal M} \times {\cal N}$ matrices respectively.

\end{itemize}

\subsection{Dressing: the Lax pairs}

\noindent Our aim is to construct the Lax pair via the dressing scheme. To achieve this we consider 
the fundamental Darboux transform (\ref{DAL1}).
Assuming specific forms for the linear time components (zero fields) of the Lax pairs of the hierarchy 
leads to two distinct models: 1) the matrix AL model, 
2) a generalization of the non-linear network equations \cite{Darboux}.

Before we proceed with the dressing it would be useful to obtain the associated Darboux-BT relations. 
From relations (\ref{BT1}) for (\ref{AL}) and (\ref{DAL1})
we obtain the set of constraints:
\begin{eqnarray}
&& B_{n+1} = Q^{-1} \big ( -\hat b_n + A_{n+1}\hat b_n^{(0)}\big ), ~~~~B_{n} = Q\big ( \hat b_n^{(0)} - \hat b_n D_{n}\big )\nonumber\\
&&  C_{n+1} =Q \Big ( b_n- D_{n+1} b_n^{(0)}\Big ), ~~~~~C_n =Q^{-1} \Big ( -b_n^{(0)}+ b_n A_{n}\Big ) \nonumber\\
&&  A_{n+1} - A_n = Q\Big ( B_{n+1} b_{n}^{(0)} -\hat b_n C_n\Big ) \nonumber\\
&&  D_{n+1} -D_n =Q^{-1} \Big (b_nB_n - C_{n+1} \hat b_n^{(0)} \Big ). \label{BTB}
\end{eqnarray}
The relations above will be used below for $\hat b^{(0)} = b^{(0)} =0 $ to obtain the hierarchy via dressing, 
and also in the next subsection to derive solutions.

Let us consider the following general form for the $V^{(\alpha),0}$ operators:
\begin{equation}
V^{(\alpha),0} = z^{\alpha}\Sigma^+ - z^{-\alpha} \Sigma^-, ~~~~\Sigma^+ = \lb \begin{matrix}
		 {\mathbb I}^{({\cal N})}&0\\
		0 & 0\end{matrix} \rb,~~~~~\Sigma^- =  \lb \begin{matrix}
		0&0\\
		0 & {\mathbb I}^{({\cal M})}
	\end{matrix} \rb, 
	 \ \label{v0}
\end{equation}
we could have considered in general $V^{(\alpha),0} = z^{\alpha}\Sigma^+ - \kappa z^{-\alpha} \Sigma^-$. Let the dressed operators be of the form
\begin{equation}
V_n^{(\alpha)} = z^{\alpha}\Sigma^+ - z^{-\alpha} \Sigma^- + \sum_{k=-(\alpha-1)}^{\alpha-1} z^{k} W_{n,k}^{(\alpha)}.
\end{equation}
We focus on the derivation of the first few members of the hierarchy. The first observation emerging from our calculations is 
that the even powers in the spectral parameter expansion provide inconsistent results, i.e the only pairs that exist are the ones for odd powers.  
This remark is in agreement with the conserved quantities emerging from the transfer matrix, indeed from the expansion we only have 
conserved quantities for even powers of $z$ or $z^{-1}$. For the first two members of the hierarchy in particular we obtain
\begin{eqnarray}
&& 
V^{(0)} = \lb \begin{matrix}
		 {\mathbb I}^{({\cal N})}&0\\
		0 & - {\mathbb I}^{({\cal M})}\end{matrix} \rb \nonumber\\
&& 
V_n^{(2)}= \lb \begin{matrix}
		 z^2{\mathbb I}^{({\cal N})} -\hat b_{n} b_{n-1} & z \hat b_{n} - z^{-1} \hat b_{n-1}\\
		z b_{n-1} - z^{-1} b_{n} &  - z^{-2} {\mathbb I}^{({\cal M})} + b_{n} \hat b_{n-1}\end{matrix} \rb.
\end{eqnarray}
It is convenient to define $\hat V^{(2)} =V^{(2)} - V^{(0)} $; this corresponds to the addition of a 
term $\sum_n \log(1-b_{n}\hat b_n)$ in the Hamiltonian (\ref{HAL}).
The semi-discrete zero curvature condition for the pair $\big (L,\ \hat V^{(2)} \big )$ lead to the equations of motion for
the generalized AL model in a matrix form:
\begin{eqnarray}
&&  \dot {\hat b}_n = \hat b_{n+1} + \hat b_{n-1} -2 \hat b_n- \hat b_n  b_n \hat b_{n-1} - \hat b_{n+1} b_n  \hat b_n \nonumber\\
&&   \dot b_n = - b_{n+1} - b_{n-1}  + 2 b_n + b_{n+1} \hat b_n b_n + b_n \hat b_n b_{n-1}. \label{GAL}
\end{eqnarray}

\noindent {\bf Remark 3.}\\
By assuming the following forms for the $V$-operators:
\begin{equation}
V^{(\alpha),0 } = z^{\alpha}\Sigma^+ + z^{-\alpha} \Sigma^- + , ~~~~
V_n^{(\alpha)} = z^{\alpha}\Sigma^+ + z^{-\alpha} \Sigma^- + \sum_{k=-(\alpha-1)}^{\alpha-1} z^{k} W_{n,k}^{(\alpha)},
\end{equation}
we arrive at a variation of the AL model, the matrix non-linear network equations \cite{Darboux}. 
Indeed, we find that $V^{(0) } = \Sigma^+  +\Sigma^- $, and 
\begin{eqnarray}
V_n^{(2)}= \lb \begin{matrix}
		 z^2{\mathbb I}^{({\cal N})} -\hat b_{n} b_{n-1} & z \hat b_{n} +z^{-1} \hat b_{n-1}\\
		z b_{n-1} + z^{-1} b_{n} &  z^{-2} {\mathbb I}^{({\cal M})} - b_{n} \hat b_{n-1}\end{matrix} \rb
\end{eqnarray}
and the corresponding equations of motion read as:
\begin{eqnarray}
&&  \dot {\hat b}_n = \hat b_{n+1} - \hat b_{n-1} +\hat b_n  b_n \hat b_{n-1} - \hat b_{n+1} b_n  \hat b_n \nonumber\\
&&   \dot b_n = b_{n+1} - b_{n-1} -b_{n+1} \hat b_n b_n + b_n \hat b_n b_{n-1},
\end{eqnarray}
which allows the cases $\hat b = b$  (${\cal N} = {\cal M}$) ; $\hat b =1$, then the latter can be thought of as a 
discretization of the generalized  (m)KdV equation. $\square$

\subsection{Solutions}

\noindent To find the one-soliton solution we consider the $b_n^{(0)} = \hat b_n^{(0)} =0$, 
also recall $C_n =Q b_{n-1},\ B_n = -Q^{-1} \hat b_{n-1} $ 
then the B\"acklund transformation (BT) relations (\ref{BTB}) reduce to
\begin{eqnarray}
&& \hat b_{n-1} = Q^{2} \hat b_nD_n, ~~~~~b_{n-1} = Q^{-2} b_n A_n  \label{AL1}\\
&& A_{n+1} -A_n =- Q^2 \hat b_n b_{n-1}, ~~~~~D_{n+1} - D_n =-Q^{-2} b_n \hat b_{n-1}.\label{AL2}
\end{eqnarray}
Suitably combining (\ref{AL1}), (\ref{AL2}):
\begin{equation}
A_{n+1} -A_n = -\hat b_n b_n A_n, ~~~~~D_{n+1} -D_n = -b_n \hat b_n D_n.
\end{equation}

To solve the matrix relations above we implement, as in the preceding section, the following ansatz for the fields:
\begin{equation}
\hat b_n ={ \hat {\mathrm b}}_n \hat {\mathrm B}, ~~~~~b_n ={\mathrm b}_n {\mathrm B}, ~~~~
A_n = {\mathbb I} + {\mathrm a}_n \hat {\mathrm B}{\mathrm B}, ~~~~~
D_n = {\mathbb I} + {\mathrm d}_n  {\mathrm B} \hat {\mathrm B},  \label{ansatz2}
\end{equation}
where $\hat {\mathrm B}$, ${\mathrm B}$ are ${\cal N}\times {\cal M}$ and ${\cal M} \times {\cal N}$ matrices respectively  that obey (\ref{TL}).

Then (\ref{AL1}), (\ref{ansatz2}) lead to:
\begin{equation}
{\hat {\mathrm b}}_n = ( Q^2)^{-(n-1)} \prod_{m=2}^n (1 +\kappa {\mathrm d}_m)^{-1}\  \hat {\mathrm b}_1, ~~~~~ 
{\mathrm b}_n = (Q^2)^{n-1} \prod_{m=2}^n (1+ \kappa {\mathrm a}_m)^{-1}\  {\mathrm b}_1.
\end{equation}
The knowledge of ${\mathrm a}_n,\ {\mathrm d}_n$ allows then the explicit computation of ${\mathrm b}_n, \hat {\mathrm b}_n$.

\subsection{The deformed oscillator-type Darboux }

\noindent From the fundamental  relation (\ref{BT1}) we obtain the corresponding Darboux-BT relations for (\ref{DAL})
\begin{eqnarray}
&&  B_n = Q \hat b_n^{(0)}, ~~~~B_{n+1} - Q^{-1}A_{n+1} \hat b_n^{(0)} = -Q^{-1} \hat b_n \nonumber\\
&&  C_{n+1} = Qb_n, ~~~~C_n - Q^{-1} b_n A_n =-Q^{-1} b_n^{(0)} \nonumber\\
&&  A_{n+1} -A_n = Q\big (B_{n+1} b_n^{(0)}  -\hat b_n C_n \big ). \label{DALq}
\end{eqnarray}
We choose to consider the following, $b_n^{(0)}=0$, then from (\ref{GAL}) it is clear that $\hat b^{(0)}_n$
satisfies the discrete heat equation 
\begin{equation}
\dot {\hat b}_{n}^{(0)} =\hat  b^{(0)}_{n+1} -2 \hat  b^{(0)}_{n}+\hat  b^{(0)}_{n-1}. \label{dheat2}
\end{equation}
Notice that this is a symmetric discretization as opposed to the forward discretization in the previous section (\ref{discr1}).
The general solution of symmetric discrete heat equation (\ref{dheat2}) is given by
\begin{equation}
\hat b_n^{(0)} = \sum_{s=1}^S c_s \xi_{s}^{n-1} e^{\Lambda_s t }, ~~~~\mbox{and/or} ~~~~\hat b_n^{(0)} = 
\int_{|\xi| =1} d\xi\  c(\xi)\xi^{n-1}e^{ \Lambda_{\xi} t }
\end{equation}
and the dispersion relation is obtained via (\ref{dheat2}): 
\begin{equation}
\Lambda_s = (\xi_s^{1 \over 2} - \xi_s^{-{1 \over 2}})^2. \label{disp22}
\end{equation}

Setting $b_n^{(0)} = 0$ in (\ref{DALq}) we end up to a set of simple equations
\begin{eqnarray}
&& Q\hat b^{(0)}_{n+1} - Q^{-1}A_{n+1} \hat b_n^{(0)} = -Q^{-1} \hat b_n \nonumber\\
&& Qb_{n-1} =  Q^{-1} b_n A_n  \nonumber\\
&& A_{n+1} -A_n =- Q^2 \hat b_n b_{n-1}.  \label{DAL0}
\end{eqnarray}

By imposing the extra constraint: $A_n = \kappa \hat b_n^{(0)} b_{n-1}+ \zeta$ ($\kappa,\  \zeta$ constants, 
see also \cite{DoikouFindlay1}) we can explicitly solve equations (\ref{DAL0}) and express the fields in terms 
of the solutions $\hat b_n^{(0)}$ of the discrete heat equation as in subsection 2.3.2. However, the solution of these equations
is not as straightforward as the ones in subsection 2.3.2.

Multi-soliton solutions can be also obtained via Bianchi's permutability theorem as in the previous section, 
however the computation is technically more demanding 
in this case. In general, the derivation of explicit expressions of solitonic as well as more general solutions 
via the fundamental Darboux transforms is more involved in this case compared to the derivation of the 
previous section for the generalized DNLS model. 
This is in fact one of the advantages of the DNLS model compared to the AL. In the subsequent 
section we are deriving  a unifying frame by means of the
discrete version of the Gelfand-Levian-Marchenko equation and appropriate linearizations in order 
to identify general solutions for both discretization schemes.

\section{The discrete GLM equation}

\noindent  The main objective in this section is to construct and solve the discrete GLM equation and 
derive general solutions for the discrete ZS-AKNS hierarchy \cite{ZakharovShabat, Ablo}.
We consider below two discretization schemes, the forward/backward corresponding to the DNLS hierarchy,
and the symmetric corresponding to the AL model 
(see also \cite{Ablo} on solutions of the generalized AL model via the inverse scattering transform).

Let us  define
\begin{equation}
{\mathrm D} = \sum_{j=-N}^N \big ( e_{jj+1} - e_{jj} \big ), 
~~~~{\mathrm D}^* = \sum_{j=-N}^N \big ( e_{j+1j} - e_{jj} \big ), \label{difference}
\end{equation}
where $e_{ij}$ are $(2N+1)\times (2N+1)$ matrices with entries: $(e_{ij})_{kl} = \delta_{ik} \delta_{jl}$,
and also introduce the linear operators associated to the two distinct discretization schemes:
\begin{enumerate}
\item{\tt Forward/backward scheme}
\begin{equation}
{\mathbb D}^{(\alpha)} = \partial_{t_{\alpha}} - {\cal D}^{\alpha}, ~~~~\alpha \geq 1, \label{linearops}
\end{equation}
where we define:
$ {\cal D} =  \lb \begin{matrix}
		 {\mathbb I}^{({\cal N})} {\mathrm D}   &0\\
		0 & w {\mathbb I}^{({\cal M})}{\mathrm D}^* \end{matrix} \rb$.

\item{\tt Symmetric scheme}

\begin{equation}
{\mathbb D}^{(\alpha)} = \partial_{t_{\alpha}} - \hat {\cal D}^{2\alpha}, ~~~~\alpha \geq 1, \label{linearops2}
\end{equation}
where $\hat {\cal D}^2 = - {\mathrm D} {\mathrm D}^*\lb \begin{matrix}
		 {\mathbb I}^{({\cal N})}  &0\\
		0 & w {\mathbb I}^{({\cal M})} \end{matrix} \rb  $.
\end{enumerate}

Let also
\begin{equation}
{\mathbb I} + F = {\mathbb I} +\lb \begin{matrix}
		0  & \hat f_{{\cal N} \times {\cal M}}\\
		f_{{\cal M} \times {\cal N}} & 0 \end{matrix} \rb
\end{equation}
 be a solution of the linear problem i.e.:
\begin{equation}
{\mathbb D}^{({\alpha})} F = 0, \label{linearproblem}
\end{equation}
and in addition require that 
${\mathrm f }\in \{ f,\ \hat f$\} are Hankel operators, i.e. ${\mathrm f}_{ij} = {\mathrm f}_{i+j}$.
Moreover, the upper lower Borel decomposition for $F$ is imposed:
\begin{equation}
\big ({\mathbb I} + K^+\big )  \big ({\mathbb I} + F \big ) = \big({\mathbb I} + K^- \big )  \label{factor}
\end{equation}
where $K^{+},\  K^{-}$ are upper, lower triangular matrices respectively. The factorization condition (\ref{factor})  
leads to the discrete GLM equation.
We drop henceforth the $^+$ in $K^+$ for simplicity, and via the factorization condition we 
obtain the discrete GLM equation, in the component form:
\begin{eqnarray}
K_{ij} + F_{ij} +\sum_{l\geq i}^NK_{il} F_{lj} =0, ~~~~ j\geq i, \label{DGLM}
\end{eqnarray}
where 
\begin{equation}
K_{ij} = \lb \begin{matrix}
		({\mathrm A}_{ij})_{{\cal N} \times {\cal N}}  & ({\mathrm B}_{ij})_{{\cal N} \times {\cal M}} \\
		({\mathrm C}_{ij})_{{\cal M} \times {\cal N}}  & ({\mathrm D}_{ij})_{{\cal M} \times {\cal M}}  \end{matrix} \rb, 
~~~~~F_{ij}= \lb \begin{matrix}
		0  & (\hat f_{ij})_{{\cal N} \times {\cal M}}\\
		(f_{ij})_{{\cal M} \times {\cal N}} & 0 \end{matrix} \rb
\end{equation}
and also $~K_{ij}^- = F_{ij} + \sum_{l\geq i}^N K_{il}F_{lj}, ~~~j\leq i$.

From the discrete GLM equation two independent sets of algebraic equations are obtained, 
in analogy to the continuous case \cite{DoFiSk1}:
\begin{eqnarray}
&& {\mathrm A}_{ij} + \sum_{l\geq i} {\mathrm B}_{il}f_{lj} =0, ~~~~
{\mathrm B}_{ij} + \hat f_{ij} + \sum_{l\geq i} {\mathrm A}_{il}\hat f_{lj} =0  \label{set1}\\ 
&& {\mathrm D}_{ij} + \sum_{l\geq i} {\mathrm C}_{il} \hat f_{lj} =0, ~~~~~ 
{\mathrm C}_{ij} + f_{ij} + \sum_{l\geq i} {\mathrm D}_{il} f_{lj} =0. \label{set2}
\end{eqnarray}
The solution of the sets of equations above leads to the simple formulas for the fields expressed solely in terms 
of the solutions of the linear problem:
\begin{eqnarray}
{\mathrm B}_{ij} + \hat f_{ij} -\sum_{l\geq i} \sum_{l'\geq i}{\mathrm B}_{il'}f_{l'l} \hat f_{lj} =0, ~~~~~
{\mathrm C}_{ij} + f_{ij} - \sum_{l\geq i} \sum_{l'\geq i}{\mathrm C}_{il'}\hat f_{l'l} f_{lj} =0,\label{solution1}
\end{eqnarray}
and in a compact form (see also e.g. \cite{DoFiSk1} for the continuous analogue)
\begin{equation}
{\mathrm B} = - \hat f \cdot \big (\mbox{id} - f \cdot \hat f \big )^{-1}, ~~~~~
 {\mathrm C}=  - f \cdot \big (\mbox{id} - \hat f \cdot  f \big )^{-1}. \label{solution2}
\end{equation}
where $\cdot$ denotes matrix multiplication according to (\ref{set1})--(\ref{solution1}).

\subsection{Discrete Calculus $\&$ Solutions}

\noindent Before we discuss the solutions of the GLM equation (\ref{solution2}), 
we should first focus on the solution of the linear problem (\ref{linearproblem}), 
which also provides the corresponding dispersion relations for each time flow.
Let us first  derive some preliminary results necessary for the solution of the linear problem.
First, the powers of the difference operators ${\mathrm D},\ {\mathrm D}^*$  
(\ref{difference}) are expressed in a compact form as:
\begin{equation}
{\mathrm D}^{\alpha} =\sum_{j=-N}^N \sum_{k=0}^{\alpha}(-1)^{\alpha -k} \lb  \begin{matrix}
		\alpha \\
		k
	\end{matrix} \rb e_{j j+k}, 
~~~~~{\mathrm D}^{*\alpha}=\sum_{j=-N}^N \sum_{k=0}^{\alpha}(-1)^{\alpha -k} \lb  \begin{matrix}
		\alpha \\
		k
	\end{matrix} \rb e_{j+k j}.
\end{equation}
Then one immediately obtains for any matrix ${\mathfrak f}= \sum_{i, j} {\mathfrak f}_{ij} e_{ij}$:
\begin{eqnarray}
&& \big ( {\mathrm D}^{\alpha} {\mathfrak f}\big )_{ij} = \sum_{k=0}^{\alpha} (-1)^{\alpha-k} \lb  \begin{matrix}
		\alpha \\
		k
	\end{matrix} \rb {\mathfrak f}_{i+k j}, ~~~\big ( {\mathfrak f}{\mathrm D}^{* \alpha} \big )_{ij} =
\sum_{k=0}^{\alpha} (-1)^{\alpha-k} \lb  \begin{matrix}
		\alpha \\
		k
	\end{matrix} \rb {\mathfrak f}_{i j+k} \nonumber\\
&& \big ( {\mathrm D}^{*\alpha} {\mathfrak f}\big )_{ij} = \sum_{k=0}^{\alpha} (-1)^{\alpha-k} \lb  \begin{matrix}
		\alpha \\
		k
	\end{matrix} \rb {\mathfrak f}_{i-k j}, ~~~\big ( {\mathfrak f}{\mathrm D}^{\alpha} \big )_{ij} =
 \sum_{k=0}^{\alpha} (-1)^{\alpha-k} \lb  \begin{matrix}
		\alpha \\
		k
	\end{matrix} \rb {\mathfrak f}_{i j-k}. \nonumber\\
\end{eqnarray}

From the linear problem (\ref{linearproblem}) and (\ref{linearops}) , (\ref{linearops2})
 we obtain the following fundamental relations:

\begin{enumerate}
\item{\tt Forward/backward}
\begin{equation}
\partial_{t_{\alpha}} \hat f= {\mathrm D}^{\alpha} \hat f,
~~~~\partial_{t_{\alpha}} f= w^{\alpha}{\mathrm D}^{*\alpha}  f,  \label{discrtime}
\end{equation}
which  lead to:
\begin{eqnarray}
&& \partial_{t_{\alpha}} \hat f_{i+j} =  \sum_{k=0}^{\alpha} (-1)^{\alpha-k} \lb  \begin{matrix}
		\alpha \\
		k
	\end{matrix} \rb \hat f_{i+k+ j}  \nonumber\\
&& \partial_{t_{\alpha}} f_{i+j} =\sum_{k=0}^{\alpha} (-1)^{\alpha-k} \lb  \begin{matrix}
		\alpha \\
		k
	\end{matrix} \rb w^{\alpha}f_{i+j-k}. \label{basic2b}
\end{eqnarray}
Note that the equations above are basically linearizations of the discetere NLS-type hierarchy (see for instance 
 (\ref{E11}), (\ref{E12}), (\ref{E21}), (\ref{E22})), up to a time rescaling. 

\item{\tt Symmetric}

Focus on the $a =1$ case associated to the AL model:
\begin{eqnarray}
&& \partial_t \hat f  =- {\mathrm D} {\mathrm D}^* \hat f, ~~~~~
\partial_t f = -w {\mathrm D} {\mathrm D}^* f , 
\end{eqnarray}
where the latter equations lead to
\begin{eqnarray}
&& \partial_{t} \hat f_{i+j} = \hat f_{i+j+1} -2 \hat f_{i+j} + \hat f_{i+j-1}  \nonumber\\
&&  \partial_{t}  f_{i+j} = w \big ( f_{i+j+1} -2 f_{i+j} + f_{i+j-1} \big ). \label{symmb}
\end{eqnarray}
Note that we have assumed $N$ to be large enough so that the fields at $\pm N$ essentially vanish.

\end{enumerate}

The general form of the solutions of the linear problem (\ref{basic2b}) are given as
\begin{eqnarray}
&& \hat f_{kj}(t) = \sum_{s =1}^S \hat b_s e^{-\hat \lambda_s( k +  j )+ \hat \Lambda^{({\alpha})}_s t_{\alpha}} + 
\int\ d\hat \lambda\  \hat b(\hat \lambda)  e^{{\mathrm i} \hat \lambda (k + j )+ 
 \hat \Lambda^{({\alpha})} t_{\alpha}}\nonumber\\
&&  f_{kj}(t) = \sum_{s =1}^S  b_s e^{ - \lambda_s (j+k) +   \Lambda^{({\alpha})}_s t_{\alpha}} + 
\int\ d\lambda\   b( \lambda)  e^{{\mathrm  i} \lambda (j +k)+ \Lambda^{({\alpha})} t_{\alpha}}, \label{solutionlinear}
\end{eqnarray}
where $S$ is an integer, the number of solitons. From expressions (\ref{solutionlinear}) 
and (\ref{linearproblem}) we obtain the dispersion relations (for the discrete part):
\begin{enumerate}
\item{\tt Forward/backward}\\
From equations (\ref{basic2b}) and (\ref{solutionlinear}) we conclude
\begin{eqnarray}
\hat \Lambda_s^{({\alpha})} = \big ( e^{-\hat \lambda_s}-1\big )^{\alpha}, ~~~~ 
\Lambda_s^{({\alpha})} =  w^{\alpha}  \big ( e^{ \lambda_s}-1 \big )^{\alpha}, \label{dispersiona} 
\end{eqnarray}
see also (\ref{disp1}), (\ref{disp2}).
\item {\tt Symmetric}\\
From equations (\ref{symmb}),  (\ref{solutionlinear}) we obtain
\begin{eqnarray}
\hat \Lambda_s^{({1})} = \big ( e^{-\hat \lambda_s\over 2}- e^{\hat \lambda_s\over 2}\big )^2, ~~~~ 
\Lambda_s^{({1})} = w\big ( e^{ \lambda_s\over 2 }- e^{ -\lambda_s\over 2 } \big )^{2}, \label{dispersion} 
\end{eqnarray}
see also (\ref{disp22}).
\end{enumerate}
Similarly, for the continuous part: $o_s \to -{\mathrm i} o,\  o\in \{\lambda,\  \hat \lambda  \}$. 
This is in fact the scheme employed in \cite{Ablo} for the discrete matrix AL model.

Let us now focus on deriving the one soliton solution, i.e. we only keep the discrete part 
of the solutions (\ref{solutionlinear}) for $S=1$, i.e.
\begin{equation}
 \hat f_{kj}(t) = \hat b e^{-\hat \lambda (j+k ) +  \hat \Lambda^{({\alpha})} t_{{\alpha}}}, ~~~~ 
f_{kj}(t) = b  e^{-\lambda (j+k) +   \Lambda^{({\alpha})} t_{\alpha}}. \label{solutionlinearb}
\end{equation}
Due to the form of the solutions (\ref{solutionlinearb}) as well as (\ref{solution1}) we consider the following ansatz for the matrix fields:
\begin{equation}
{\mathrm B}_{kj} = \hat L_k(t_{\alpha}) e^{- \hat \lambda j },~~~~~{\mathrm C}_{kj} = L_k(t_{\alpha}) e^{- \lambda j } \label{fields2}
\end{equation}
Substituting (\ref{solutionlinearb}) and (\ref{fields2}) into (\ref{solution1}) we conclude, after having assumed the 
typical relations: $\hat b b \hat b = \kappa \hat b,\  b \hat b b =\kappa  b$,
\begin{eqnarray}
{\mathrm B}_{kj}(t) = -{e^{-\hat\lambda( k+j) +\hat \Lambda^{\alpha}t_{\alpha}}\over1-\kappa {\mathrm h}_k(t)}\ \hat b, ~~~~~~~~
{\mathrm C}_{kj}(t) = -{e^{-\lambda (k+j) + \Lambda^{\alpha}t_{\alpha}}\over1-\kappa {\mathrm h}_k(t)}\  b, \label{bc}
\end{eqnarray}
where ${\mathrm h}_k(t) = {e^{-2(\lambda + \hat \lambda)k} e^{ (\Lambda^{\alpha} + \hat \Lambda^{\alpha})t_{\alpha}}
\over(e^{-(\lambda+ \hat \lambda)} -1)^2}$,  
having assumed vanishing boundary conditions at $k=N$. In the scaling limit $N\to \infty,\ \delta \sim {1\over N} \to 0$: $\lambda \to \delta \lambda$ 
and $e^{\lambda k} \to e^{\lambda x}$ and $e^{\lambda} \to 1 + \lambda$. Comparison between ``local'' elements of ${\mathrm B},\  {\mathrm C}$ (\ref{bc})
and (\ref{xy2}) shows coincidence of the corresponding expressions.

\section{Comments}

\noindent Solutions of the non-linear integrable ODEs of the hierarchies under study can be expressed in terms of 
suitable local elements of ${\mathrm B},\  {\mathrm C}$ (\ref{solution1}) (see also \cite{Ablo}).
The latter statement is naturally the discrete analogue of the continuum case, where solutions of 
the non-linear integrable PDEs turn out to be ``diagonal''  
solutions of the continuous GLM equation (see e.g. \cite{ZakharovShabat, Ablo2, DoFiSk1, Pope}). 
In this context the pertinent issue, which will be fully addressed elsewhere, is the systematic derivation of solutions 
of the non-linear integrable ODEs in terms of local elements of ${\mathrm B},\  {\mathrm C}$ via the discrete analogue of the  
``dressing'' scheme as described in \cite{Ablo2} and \cite{Pope} (see also recent generalizations on local and non-local PDEs in 
\cite{BDMS, DMSW}). The derivation of general  solutions of the non-linear ODEs given any initial profile
for the linear solutions is then possible. 

The same applies to the simple, but surprisingly general novel expressions (\ref{lin1})-(\ref{y}) (see also comments below, 
and the simple solutions presented as examples)
emerging from the local Darboux transform (\ref{typeI}). We obtain similar results in subsection 3.3 for the AL scheme (\ref{dheat2})-(\ref{DAL0}).
We also extend this generic result in the continuum NLS case in Remark 1 (see in particular 
equations (\ref{contgen}) and (\ref{heat}) in terms of e.g. the heat kernel, as well as comments below). 
An interesting consequence of that is the discretization 
of the Burgers equation presented in Remark 2.
We should once more emphasize that these expressions provide part of the most significant findings 
of this investigation together with the  forward/backwrard  linearization and discretization schemes and the discrete calculus
 introduced in section 4  for the DNLS model via the use of the underlying Grassmannian structure, 
and are very much  in the spirit of the dual description presented in \cite{Sklyanin}.
These are all novel expressions even in the scalar case as far as we know.

We have only considered here periodic or vanishing boundary conditions
for the solutions of the integrable ODEs. The significant point then is the  implementation of 
integrable space and time integrable  boundary conditions \cite{Sklyanin2, DoFiSk2, CauCra} in 
the discrete systems, and the effect of these boundary conditions on the
behavior of the solutions. Having systematically studied the semi discrete version of the AKNS scheme the next natural step is to 
consider the full discrete space and time picture \cite{ABS, HJN},  along the lines described in \cite{Schiff}, 
but also from the algebraic/Hamiltonian perspective. The fully discrete case represents various technical and conceptual difficulties,
that are primarily associated with the consistent simultaneous  discretizations of both space and time directions in such 
a way that integrabilty is ensured. These are all intriguing open questions, that  will be systematically addressed  in future works.

Finally a very interesting direction to pursue is the use of finite set theoretic solutions of the Yang--Baxter equation (YBE), 
or Yang-Baxter maps  \cite{Ves, Pap}
in the context of multi soliton interactions
when constructing the soliton lattice via Bianchi's permutability (see also relevant explicit results in subsection 2.4 particularly part 2), 
but also the use of infinite set theoretic solutions in relation to global transforms (integral transforms in the continuous case)
of section 4 and the GLM equation. These are particularly significant questions, especially in view of recent findings 
on connections between set theoretic solutions of the YBE and quantum integrability \cite{DoiSmo}.



\subsection*{Acknowledgments}

\noindent  We are indebted to C. Eilbeck, and S.J.A. Malham for illuminating discussions.
AD acknowledges support from the  EPSRC research grant  EP/R009465/1. 
SS is supported by Heriot-Watt University via a J. Watt scholarship.

\appendix

\section{Higher conserved quantities}
\noindent Expansion of the transfer matrix (\ref{transfer}) in powers of $\lambda^{-1}$ provides
non-local conserved quantities
\begin{equation}
{\mathfrak t}(\lambda) = \sum_{k}{\tau_k \over \lambda^k}.
\end{equation}
To obtain local integrals of motion we consider the expansion of the 
$ln({\mathfrak t}(\lambda))$. After some tedious but straightforward algebra  we conclude that 
the general relation that provides the 
local integrals of motion is given as:
\begin{equation}
H_k = \tau_k - \sum_{j_1 +2j_2 +...+(k-1)j_{k-1}=k}{1\over \mbox{max}(j_1! j_2!...j_{k-1}!)}H_1^{j_1} H_2^{j_2} 
\ldots H_{k-1}^{j_{k-1}}. \label{generic}
\end{equation}
The first few local integrals of motion up to $H_4$ are reported in  section 2 (\ref{eq:ALP_ExDNLSP}). 
Indeed, from (\ref{generic}) we obtain for instance:
\begin{eqnarray}
&& H_2= \tau_2 - {1\over 2} H_1^2, \nonumber\\
&& H_3 =\tau_3 - H_1 H_2 - {1\over 6} H_1^3, \nonumber\\
&& H_4 = \tau_4 - H_1 H_3  - {1\over 2} H_2^2 -{1\over 2}H_1^2 H_2 - {1\over 24} H_1^4,\nonumber\\
&& \ldots 
\end{eqnarray}
which lead to (\ref{eq:ALP_ExDNLSP}).

The $V$-operator associated to the 4th integral of motion $H_4$ is given in (\ref{fourth}), and in particular $w_{n,0}^{(3)}$ is given by
the following expressions (we provide the entries of the matrix):
\begin{eqnarray}
&&(w_{n,0}^{(3)})_{11}=  X_n Y_{n-1}{\mathbb N}_{n-1} + {\mathbb N}_{n}X_n Y_{n-1} -X_{n}Y_{n-2} -X_{n+1} Y_{n-1} \nonumber\\
&&(w_{n,0}^{(3)})_{12}= X_{n+2}- X_n Y_{n-1} X_n - {\mathbb N}_{n+1} X_{n+1} -X_{n+1} Y_n X_n -{\mathbb N}_nX_{n+1} +
 {\mathbb N}_{n}^2 X_n \nonumber\\
&&(w_{n,0}^{(3)})_{21}= Y_{n-3} - Y_{n-2}{\mathbb N}_{n-2} - Y_{n-2} {\mathbb N}_{n-1} -Y_{n-1} X_{n-1} Y_{n-2} + 
Y_{n-1}{\mathbb N}_{n-1}^2 -Y_{n-1} X_n Y_{n-1} \nonumber\\
&&(w_{n,0}^{(3)})_{22}= Y_{n-2} X_n -Y_{n-1} {\mathbb N}_{n-1} X_n +Y_{n-1} X_{n+1} - Y_{n-1} {\mathbb N}_n X_n.
\end{eqnarray}


\end{document}